\documentclass[bibyear]{aa} % for the letters
\usepackage{graphicx}
\usepackage{txfonts}
\usepackage{natbib}
\bibpunct{}{}{,}{a}{}{,} %% natbib format for A&A and ApJ
\begin{document}

%
% Personal definitions
\def\hi {H\,{\sc i}}
\def\hii {H\,{\sc ii}}
\def\water {H$_2$O}
\def\meth {CH$_{3}$OH}
\def\dg{$^{\circ}$}
\def\kms{km\,s$^{-1}$}
\def\ms{m\,s$^{-1}$}
\def\jyb{Jy\,beam$^{-1}$}
\def\mjyb{mJy\,beam$^{-1}$}
\def\solmass {\hbox{M$_{\odot}$}}
\def\solum {\hbox{L$_{\odot}$}} 
\def\d {$^{\circ}$}
\def\n {$n_{\rm{H_{2}}}$}
\def\kmsg{km\,s$^{-1}$\,G$^{-1}$}
\def\tbo {$T_{\rm{b}}\Delta\Omega    $}
\def\tb {$T_{\rm{b}}$}
\def\om{$\Delta\Omega$}
\def\dvi {$\Delta V_{\rm{i}}$}
\def\dvz {$\Delta V_{\rm{Z}}$}
\def\code {FRTM code}
\title{EVN observations of 6.7~GHz methanol maser polarization in massive star-forming regions III. The flux-limited sample. }

\author{G.\ Surcis  \inst{1}
  \and 
  W.H.T. \ Vlemmings \inst{2}
 \and
  H.J. \ van Langevelde \inst{1,3}
  \and
  B. \ Hutawarakorn Kramer \inst{4,5}
  \and
  A. Bartkiewicz \inst{6}
  \and
  M.G. \ Blasi \inst{7}
  }

\institute{Joint Institute for VLBI in Europe, Postbus 2, 7990 AA Dwingeloo, The Netherlands
 \email{surcis@jive.nl}
 \and
 Chalmers University of Technology, Onsala Space obssectervatory, SE-439 92 Onsala, Sweden
 \and
 Sterrewacht Leiden, Leiden University, Postbus 9513, 2300 RA Leiden, The Netherlands
 \and
 Max-Planck Institut f\"{u}r Radioastronomie, Auf dem H\"{u}gel 69, 53121 Bonn, Germany
 \and
 National Astronomical Research Institute of Thailand, Ministry of Science and Technology, Rama VI Rd., Bangkok 10400, Thailand
 \and
 Centre for Astronomy, Faculty of Physics, Astronomy and Informatics, Nicolaus Copernicus University, Grudziadzka 5, 87-100 Torun, Poland
 \and
 Università degli Studi della Basilicata, Viale dell'Ateneo Lucano 10, 85100, Potenza, Italy
  }

\date{Received ; accepted}
\abstract
%context heading (optional)
{Theoretical simulations and observations at different angular resolutions have shown that magnetic fields have a central role in
massive star formation. Like in low-mass star formation, the magnetic field in massive young stellar objects can either 
be oriented along the outflow axis or randomly.}
{Measuring the magnetic field at milliarcsecond resolution (10-100~au) around a substantial number of massive 
young stellar objects permits determining with a high statistical significance whether the direction of the
magnetic field is correlated with the orientation of the outflow axis or not. } 
{In late 2012, we started a large VLBI campaign with the European VLBI Network to measure the linearly and circularly 
polarized emission of 6.7 GHz 
\meth ~masers around a sample of massive star-forming regions. This paper focuses on the first seven observed sources, 
G24.78+0.08, G25.65+1.05, G29.86-0.04, G35.03+0.35, G37.43+1.51, G174.20-0.08, and G213.70-12.6. For all these sources, molecular 
outflows have been detected in the past.}
{We detected a total of 176 \meth ~masing cloudlets toward the seven massive star-forming regions, 19\% of which show linearly
polarized emission. The \meth ~masers around the massive young stellar object MM1 in G174.20-0.08 show neither linearly nor 
circularly polarized emission. The linear polarization vectors are well ordered in all the other massive young stellar objects. We 
measured significant Zeeman splitting toward both A1 and A2 in G24.78+0.08, and toward G29.86-0.04 and G213.70-12.6.}
{By considering all the 19 massive young stellar objects reported in the literature for which both the orientation of the magnetic 
field at milliarcsecond resolution and the orientation of outflow axes are known, we find evidence that the magnetic field 
(on scales 10-100~au) is preferentially oriented along the outflow axes. }
\keywords{Stars: formation - masers: methanol - polarization - magnetic fields - ISM: individual: 
G24.78+0.08, G25.65+1.05, G29.86-0.04, G35.03+0.35, G37.43+1.51, G174.20-0.08, G213.70-12.6. }
\titlerunning{Magnetic field and outflows: the flux-limited sample.}
\authorrunning{Surcis et al.}

\maketitle
%________________________________________________________________
\section{Introduction}
\label{intro}
The \textit{\textup{core accretion}} model describes the formation of high-mass stars as a scaled-up version of the formation process of 
low-mass stars (e.g., McKee \& Tan \cite{mck03}). Specifically, it is proposed that massive stars form through gravitational 
collapse, which involves disk-assisted accretion to overcome radiation pressure and matter-ejection perpendicular to the disk to 
redistribute the angular momentum (e.g., McKee \& Tan \cite{mck03}). Nevertheless, only when the magnetic field has been taken into 
consideration, the theoretical simulations begin to faithfully reproduce the observations (e.g., Peters et al. \cite{pet11}; 
Seifried et al. \cite{sei12a}; Myers et al. \cite{mye13}). This suggests that magnetic fields might play a role in 
massive star formation just as importantly as in the formation of low-mass stars. In low-mass star formation the 
magnetic field is thought to slow the collapse, to transfer the angular momentum, and to power the outflow (e.g., McKee \& Ostriker 
\cite{mck07}). However, in studies on low-mass star formation it is still an open debate whether the magnetic field aligns with the 
molecular outflows. Recently, two independent polarization surveys of low-mass protostellar cores, 
which were carried out at different spatial resolutions, showed two contrasting results. Hull et al. (\cite{hul13}) found
on scales of a few 100 to 1000~au no correlation between magnetic field orientation and outflow axis in low-mass young 
stellar objects (YSOs), while Chapman et al. (\cite{cha13}) found a good alignment on larger scales ($>2\times10^3$~au). 
Similar conflicting results were also 
found toward massive YSOs (Surcis et al. \cite{sur12, sur13}; hereafter Paper~I and Paper~II, respectively; Zhang et al. 
\cite{zha14}). Surcis and collaborators started a VLBI-observations campaign of 6.7 GHz \meth ~masers toward massive 
star-forming regions (SFRs) to determine if there exists any correlation between the orientations of the magnetic field and of the 
outflow on very small scales (tens of au). Based on the results of nine sources, they found evidence that on scales of 
10-100~au the magnetic field around massive YSOs is preferentially oriented along the outflow (Paper~II). On the 
other hand, based on a larger sample (21~sources), Zhang et al. (\cite{zha14}) reported that at arcsecond resolution 
(thousands of au) the outflow axis appears to be randomly oriented with respect to the magnetic field in the core.
These contrasting results might be due to the different resolutions; indeed, Zhang et al. (\cite{zha14}) postulated that 
angular momentum and dynamic interactions, possibly due to close binary or multiple systems, dominate magnetic fields at
scales of about $10^3$~au.\\
\indent Before drawing any conclusion, it is important to improve the statistics by enlarging the number of 
massive SFRs toward which the orientation of the magnetic field at milliarcsecond (mas) resolution has been measured.
%
%\section{The flux-limited sample}
%\label{SEVNG}
Therefore, we have selected a flux-limited sample of 37 massive SFRs with declination $>-9$\d ~and a total \meth ~maser 
single-dish flux greater than 50~Jy from the 6.7 GHz \meth ~maser catalog of Pestalozzi et al. (\cite{pes05}). To increase the 
likelihood  of detecting circularly polarized \meth ~maser emission ($\leq$ 1\%) and thus allow the determination of the magnetic 
field strength, we have excluded the six regions hosting \meth ~maser that in recent single-dish 
observations showed a total flux below 20~Jy (Vlemmings et al. \cite{vle11}). The total number of massive SFRs of the flux-limited 
sample is thus 31. The polarimetric 6.7 GHz \meth ~maser observations, and the subsequent measurement of the magnetic field orientation, 
of twelve of these SFRs had already been published in the recent past (Vlemmings et 
al. \cite{vle10}; Surcis et al. \cite{sur09, sur11a, sur14}; Papers~I and II). Therefore, 19 massive SFRs remain 
to be observed. We were given European VLBI Network\footnote{The European VLBI Network is a joint facility of European, Chinese, 
South African and other radio astronomy institutes funded by their national research councils.} (EVN) time to observe all of them at 
6.7~GHz in several sessions between November 2012 and June 2015 (see Sect.~\ref{obssect}). Here, we present the results of 
the first seven observed sources. The results of the remaining twelve sources will be published in future papers of the present series 
as soon as they are observed and the data are fully analyzed.
Following the organization of Paper~I and II, the sources are briefly introduced in 
Sects.~\ref{G24_intro}--\ref{G213_intro}, while the observations and our analysis are described in Sect.~\ref{obssect}. 
The results, which are presented in Sect.~\ref{res}, are discussed in Sect.~\ref{discussion}, where we update our previous statistics.
\section{Massive star-forming regions}
\label{SEVNG}
\subsection{\object{G24.78+0.08}}
\label{G24_intro}
G24.78+0.08 is one of the most studied massive SFR (e.g., Codella et al. \cite{cod97}; 
Cesaroni et al. \cite{ces03}; Beltr\'{a}n et al. \cite{bel06,bel11}). The region is located at a kinematic distance of 7.7~kpc 
(Codella et al. \cite{cod97}) and contains four centers of star formation, named from A to D (Furuya et al. \cite{fur02}). 
Clump A is composed of two subclumps (A1 and A2) that had been resolved in five distinct cores by Beltr\'{a}n et al.
(\cite{bel11}). These cores (named A1, A1b, A1c, A2, and A2b) are aligned in a southeast-northwest direction
coincident with the CO-outflow (position angle $\rm{PA_{outflow}^{\rm{^{12}CO}}}=-40$\d) that is associated with core A2 
($M_{\rm{A2}}=22$~\solmass; Beltr\'{a}n et al. \cite{bel11}). Codella et al. (\cite{cod13}) confirmed the source embedded in
A2 as the driving source of 
the outflow by imaging the SiO emission ($\rm{PA_{outflow}^{\rm{SiO}}}\approx-45$\d) at arcsecond resolution with the SMA. Core A1 
($M_{\rm{A1}}=16$~\solmass; Beltr\'{a}n et al. 
\cite{bel11}) is associated with a hypercompact (HC) \hii ~region (Galv\'{a}n-Madrid et al. \cite{gal08}). Both A1 and A2 are 
embedded in toroids that rotate clockwise with position angle PA$_{\rm{A1}}=+50$\d ~and PA$_{\rm{A2}}=+40$\d
~(Beltr\'{a}n et al. \cite{bel04,bel05,bel11}), that is, the toroids are almost perpendicular to the axis of the CO-outflow. 
Moscadelli et al. (\cite{mos07}) detected 6.7 GHz \meth ~masers around both A1 and around A2. The \meth ~masers show velocity 
distributions consistent with the velocities of the toroidal structures ($V^{\rm{A1}}_{\rm{toroid}}=109.0-112.2$~\kms ~and 
$V^{\rm{A2}}_{\rm{toroid}}=109.6-111.6$~\kms; Beltr\'{a}n et al. \cite{bel11}). Moreover, Moscadelli et al. (\cite{mos07}) also 
detected 22 GHz \water ~masers around A1. These masers trace a fast ($\sim$40~\kms) expanding shell surrounding the HC~\hii ~region.\\
\indent Using  observations made with the Effelsberg 100 m telescope, Vlemmings et al. (\cite{vle11}) measured a Zeeman splitting of 
the \meth ~maser emission of $\Delta V_{\rm{Z}}=(+0.50\pm0.08)$~\ms.
\subsection{\object{G25.65+1.05}}
\label{G25_intro}
The massive SFR G25.65+1.05 (also known as IRAS\,18316-0602 and RAFGL7009S) is located at a kinematic distance of 3.17~kpc (Molinari
et al. \cite{mol96}). The region is associated with a weak and irregular compact radio source that was initially classified as an 
ultracompact (UC) \hii ~region (Kurtz et al. \cite{kur94}; Walsh et al. \cite{wal98}). The radio source spatially coincides with an
unresolved infrared source (Zavagno et al. \cite{zav02}; Varricatt et al. \cite{var10}) and with submillimeter emissions at 
350~$\rm{\mu}$m, 450~$\rm{\mu}$m, and 850~$\rm{\mu}$m (Hunter et al. \cite{hun00}; Walsh et al. \cite{wal03}). A bipolar 
CO-outflow ($\rm{PA}_{\rm{outflow}}^{\rm{^{12}CO}}\approx-65$\d) centered on the radio source was first detected by Shepherd 
\& Churchwell (\cite{she96}). Recently, S\'{a}nchez-Monge et al. (\cite{san13}) mapped the outflow using a more reliable jet tracer, 
 SiO emission. They detected both the red- ($+45.8$~\kms$<V_{\rm{red}}^{\rm{SiO(2-1)}}<+88.1$~\kms) and blue-shifted 
($+5.9$~\kms$<V_{\rm{blue}}^{\rm{SiO(2-1)}}<+39.5$~\kms) lobes of the jet or outflow ($\rm{PA_{outflow}^{\rm{SiO}}}=-15$\d). 
Furthermore, four 6.7 GHz \meth ~masers were detected near the continuum peak of the radio source; they are linearly distributed southward 
(Walsh et al. \cite{wal98}). The \meth ~maser velocities suggest an association with the radio source, possibly with a disk and
not with the bipolar outflow (Zavagno et al. \cite{zav02}).\\
\indent Finally, Vall\'{e}e \& Bastien (\cite{val00}) mapped the magnetic field toward the radio source at 760~$\rm{\mu}$m, finding
an orientation of the magnetic field of $\rm{\Phi_{B}^{760\mu m}=+8^{\circ}\pm16}$\d ~(scale of $10^4$~au). 
A Zeeman splitting of the 6.7 GHz \meth ~maser emission of $\Delta V_{\rm{Z}}=(+0.46\pm0.05)$~\ms ~was measured 
with the Effelsberg 100 m telescope (Vlemmings et al. \cite{vle11}).
\begin {table*}[t]
\caption []{Observational details.} 
\begin{center}
\scriptsize
\begin{tabular}{ l c c c c c c c c c c c c }
\hline
\hline
\,\,\,\,\,(1)        &(2)           & (3)              &  (4)         &  (5)         & (6)              & (7)      & (8)     & (9)   &(10)&(11)&(12)&(13)  \\
Source               & Program      & observation      & Calibrator   & Polarization & Beam size        & Position & rms     & $\sigma_{\rm{s.-n.}}$\tablefootmark{b}  & \multicolumn{4}{c}{Estimated absolute position using FRMAP} \\ 
                     & code         & date             &              &  angle       &                  & Angle    &         &       &$\alpha_{2000}$           & $\delta_{2000}$            & $\Delta\alpha$\tablefootmark{a} & $\Delta\delta$\tablefootmark{a}     \\ 
                     &              &                  &              &  (\d)        &(mas~$\times$~mas)& (\d)     & ($\frac{\rm{mJy}}{\rm{beam}}$) &  ($\frac{\rm{mJy}}{\rm{beam}}$) & ($\rm{^{h}:~^{m}:~^{s}}$) & ($\rm{^{\circ}:\,':\,''}$) &     (mas)             & (mas) \\ 
\hline
G24.78+0.08          & ES072        & 30 May 2013      & J2202+4216   & $-31\pm 4$   & $10.4\times4.0$  & -36.14   & 4       & 8     & +18:36:12.563            & -07:12:10.787              & 0.4 & 3.7 \\
G25.65+1.05          & ES072        & 31 May 2013      & J2202+4216   & $-31\pm 4$   & $11.9\times3.5$  & -39.94   & 2       & 35    & +18:34:20.900            & -05:59:42.098              & 2.3 & 18.3\\
G29.86-0.04          & ES072        & 01 June 2013     & J2202+4216   & $-31\pm 4$   & $9.0\times3.6$   & -40.71   & 4       & 6     & +18:45:59.572            & -02:45:01.573              & 8.3  & 172.5\\
G35.03+0.35          & ES069        & 04 Nov. 2012     & J2202+4216   & $-30\pm 2$   & $6.3\times4.8$   & -34.53   & 4       & 6     & +18:54:00.660            & +02:01:18.551              & 7.2 & 167.7 \\
G37.43+1.51          & ES072        & 02 June 2013     & J2202+4216   & $-31\pm 4$   & $8.3\times3.8$   & -52.43   & 3       & 22    & +18:54:14.229            & +04:41:41.138              & 7.0 & 81.2 \\
G174.20-0.08         & ES069        & 04 Nov. 2012     & J0555+3948   & $-73\pm 5$   & $7.7\times4.3$   & -28.78   & 4       & 5     & +05:30:48.020            & +33:47:54.611              & 0.7 & 1.0 \\
G213.70-12.6         & ES069        & 03 Nov. 2012     & J0555+3948   & $-73\pm 5$   & $7.4\times5.4$   & -3.13    & 4       & 10    & +06:07:47.860            & -06:22:56.626              & 2.1 & 17.9\\
\hline
\end{tabular}
\end{center}
\tablefoot{
\tablefoottext{a}{Formal errors of the fringe rate mapping.}
\tablefoottext{b}{Self-noise in the maser emission channels (e.g., Sault \cite{sau12}).}
}
\label{Obs}
\end{table*}
\subsection{\object{G29.86-0.04}}
\label{G29_intro}
G29.86-0.04 is at a kinematic distance of 7.4~kpc and has a velocity $V_{\rm{lsr}}^{\rm{C^{18}O}}=+101.85$~\kms ~(de Villiers et al. 
\cite{dev14}). Caswell et al. (\cite{cas93, cas95}) detected 12 GHz and 6.7 GHz \meth ~masers toward the region. The 6.7 GHz 
\meth ~masers show an arched distribution ($150~$mas~$\times~340$~mas) accompanied by a clear velocity gradient at mas
resolution (Fujisawa et al. \cite{fuj14}). The \meth ~masers are associated with one of the two cores that were detected toward the 
region (Hill et al. \cite{hil05, hil06}). No 22 GHz \water ~masers have been detected (Breen \& Ellingsen \cite{bre11}). A 
bipolar CO-outflow is associated with the 6.7 GHz \meth ~masers (de Villiers et al. \cite{dev14}). While the redshifted
lobe ($+104$~\kms$<V_{\rm{red}}^{\rm{^{13}CO}}<+110$~\kms) of the outflow is oriented almost south-north on the 
plane of the sky ($\rm{PA}_{\rm{red-shifted}}^{\rm{^{13}CO}}\approx+6$\d), the blueshifted lobe 
($+90$~\kms$<V_{\rm{blue}}^{\rm{^{13}CO}}<+96.5$~\kms) bends westwards passing from about 6\d ~to 
$\rm{PA}_{\rm{blue-shifted}}^{\rm{^{13}CO}}\approx+60$\d ~(de Villiers et al. \cite{dev14}).\\
\indent The 6.7 GHz \meth ~maser Zeeman-splitting was measured to be 
$\Delta V_{\rm{Z}}=(+0.50\pm0.08)$~\ms ~with the Effelsberg 100
m telescope (Vlemmings et al. \cite{vle11}).
\subsection{\object{G35.03+0.35}}
\label{G35_intro}
The extended green object (EGO) G35.03+0.35 hosts several massive YSOs at early evolutionary stages (Cyganowski et al. \cite{cyg09};
Paron et al. \cite{par12}). This massive SFR ($V_{\rm{lsr}}=+51.5$~\kms; Paron et al. \cite{par12}) is located at a kinematic distance 
of 3.43$^{+0.38}_{-0.38}$~kpc (Cyganowski et al. \cite{cyg09}). Four of the five radio continuum sources that were detected 
toward the region (CM1--5) are aligned with the bipolar morphology of the 4.5~$\mu$m emission 
($\rm{PA^{4.5~\mu m}}=+27^{\circ}$; Cyganowski et al. \cite{cyg11}). CM1, 
which is a well-known UC~\hii ~region, and CM4 are associated with the southwestern lobe of the 4.5~$\mu$m emission, CM3 is associated 
with the northeastern lobe, and CM2 is located between the two lobes. The symmetric spacing of CM3 and CM4 relative to CM2 might be the 
signature of knots in an ionized jet (Cyganowski et al. \cite{cyg11}). Furthermore, the radio spectral index of CM2 suggests that the 
radio source might either be a HC~\hii ~region or the product of an ionized wind that hits the surrounding gas 
(Cyganowski et al. \cite{cyg11}; Paron et al. \cite{par12}). Paron et al. (\cite{par12}) detected a bipolar $^{12}$CO-outflow
at a resolution of tens of arcseconds 
(beam size~=~22~ arcsec)  that is coincident in position with the whole 4.5~$\mu$m emission. The axis of 
the $^{12}$CO-outflow is oriented almost along the line of sight, with the redshifted lobe 
($+58$~\kms$<V_{\rm{red}}^{\rm{^{12}CO}}<+66$~\kms) southeast and the blueshifted lobe 
($+37$~\kms$<V_{\rm{blue}}^{\rm{^{12}CO}}<+49$~\kms) northwest
of the axis.\\
\indent \meth, ~\water, and OH masers were detected toward CM2 (Forster \& Caswell \cite{for99}; Argon et al. \cite{arg00};
Cyganowski et al. \cite{cyg09}; Pandian et al. \cite{pan11}). The 6.7 GHz \meth ~masers, which are all blueshifted,
lie on the ``waist'' between the two lobes of the 4.5~$\mu$m emission and show a complex morphology at scales of 10~mas 
(Cyganowski et al. \cite{cyg09}; Pandian et al. \cite{pan11}). Recently, Caswell et al. (\cite{cas13}) measured a persistent
linearly polarized emission of the OH masers over several years. A large Zeeman splitting of $\Delta V_{\rm{Z}}=(+1.22\pm0.23)$~\ms of the 6.7 GHz \meth ~maser emission was
measured by Vlemmings et al. (\cite{vle11}) with the Effelsberg 100 m telescope.
\subsection{\object{G37.43+1.51}}
\label{G37_intro}
The massive SFR G37.43+1.51 coincides with the IRAS source 18517+0437 ($V_{\rm{lsr}}=+44.1$~\kms; L\'{o}pez-Sepulcre et al. 
\cite{sep10}), and it is located at a parallax distance of 1.88$^{+0.08}_{-0.08}$~kpc (Wu et al. \cite{wu14}). L\'{o}pez-Sepulcre et al.
(\cite{sep10}) detected a C$^{18}$O-outflow oriented north-south ($\rm{PA_{outflow}^{\rm{C^{18}O}}}=-4$\d) with the redshifted
lobe  ($+46$~\kms$<V_{\rm{red}}^{\rm{^{18}CO}}<+50$~\kms) and the blueshifted lobe ($+38$~\kms$<V_{\rm{blue}}^{\rm{^{18}CO}}<+42$~\kms) 
located north and south, respectively. The C$^{18}$O-outflow is associated with 6.7 GHz \meth ~masers that have been detected in a 
linear distribution northwest-southeast with a clear velocity gradient (Schutte et al. \cite{sch93}; Fujisawa et al. \cite{fuj14}; 
Wu et al. \cite{wu14}). Vlemmings (\cite{vle08}) measured a Zeeman
splitting of the 6.7 
GHz \meth ~maser of $\Delta V_{\rm{Z}}=(+0.75\pm0.09)$~\ms.
\subsection{\object{G174.20-0.08}}
\label{G174_intro}
In G174.20-0.08, which is  better known as AFGL\,5142, two centers of massive star formation were identified: IRAS\,05274+3345 and 
IRAS\,05274+3345-East (Hunter et al. \cite{hun95}; Torrelles et al. \cite{tor92}). This massive SFR is located at a kinematic distance 
of 1.8~kpc (Snell et al. \cite{sne88}). IRAS\,05274+3345-East ($V_{\rm{lsr}}=-1.0$~\kms; Zhang et al. \cite{zha07}) 
hosts five 1.3~mm cores (MM--1 to MM-5) and three CO-outflows (Zhang et al. \cite{zha07}). Outflow-C is associated with core MM--1, which powers 22 GHz \water ~and 6.7 GHz \meth ~masers (Goddi et al. \cite{god07,god11}). 
While the proper-motion measurements of the \water ~masers trace the expansion of the collimated outflow-C
($\rm{PA_{outflow}^{\rm{H_{2}O}}}=-40$\d; Goddi et al. \cite{god11}), the \meth ~masers instead trace an infall of gas onto the 
central massive protostar (Goddi et al. \cite{god11}). Palau et al. (\cite{pal11}) found evidence of a possible disk perpendicular to outflow-C by observing complex organic 
molecules. No Zeeman splitting of the \meth ~maser emission was 
measured toward AFGL\,5142 with the Effelsberg 100 m telescope ($<0.08\%$; Vlemmings \cite{vle08}).\\
\indent 
\subsection{\object{G213.70-12.6}}
\label{G213_intro}
The source G213.70-12.6 ($D=0.83$~Kpc; Herbst \& Racine \cite{her76}) is better known under the name Monoceros~R2 (hereafter 
Mon~R2) and is composed of several \hii ~regions and YSOs (e.g., Howard et al. \cite{how94}; Carpenter et al. \cite{car97}; 
Preibisch et al. \cite{pre02}). G213.70-12.6 hosts several infrared sources, the brightest of which is IRS\,3 ($L=14000$~\solum;
Henning et al. \cite{hen92}). IRS\,3 is a compact cluster of massive YSOs (Preibisch et al. \cite{pre02}) that is located at 
$\sim50''\!$ northeast of the center of a giant CO-outflow ($\rm{PA}_{\rm{outflow}}^{\rm{CO}}\approx-45$\d)
that is powered by IRS\,6 (Xu et al. \cite{xu06}; beam size 15$''$). Very recently, a bipolar 
$^{12}\rm{CO(2-1)}$ outflow ($\rm{PA_{outflow}^{\rm{^{13}CO(2-1)}}}=+53$\d) has been detected toward IRS\,3 
(Dierickx et al. \cite{die15}; beam size $\sim0.5''$). Its redshifted lobe ($+21$~\kms$<V_{\rm{red}}^{\rm{^{13}CO(2-1)}}<+26$~\kms) 
is located southwest of the source and its blueshifted lobe ($-5$~\kms$<V_{\rm{blue}}^{\rm{^{13}CO(2-1)}}<+2$~\kms) lies to the northeast.
One of the massive YSOs of IRS\,3, called Star~A ($12$~\solmass$<M<15$~\solmass; Preibisch et al. \cite{pre02}), is associated with 
6.7 GHz \meth ~masers that lie in a northeast-southwest linear distribution of 170~mas ($\rm{PA_{CH_{3}OH}^{2000}}\approx40$\d; 
Minier et al. \cite{min00}). Moreover, Star~A is located along the axis of the $^{12}\rm{CO(2-1)}$ outflow (Fig.~5 of 
Dierickx et al. \cite{die15}).\\
\indent Curran \& Chrysostomou (\cite{cur07}) measured using polarimetric observations at 850~$\mu$m a magnetic field strength 
of $\sim$0.2~mG throughout G213.70-12.6. The polarization percentage around IRS\,3 decreases below 1\%, and the magnetic 
field at 
a resolution of 6$''\!\!.$18 changes its orientation from north-south to east-west (see Fig.~1 of Curran \& Chrysostomou \cite{cur07}).
At a spatial resolution of $0''\!.$97, the polarization vectors of the 2.16~$\mu$m emission form an elliptical pattern in a $\sim10''$ region around IRS\,3 ($\rm{PA}_{\rm{B-pattern}}=-40$\d; Yao et al. \cite{yao97}).
More recently, Simpson et al. (\cite{sim13}) measured the near-infrared (2~$\mu$m) polarimetry of IRS\,3 at a spatial resolution of 
0$''\!\!.$2. They found that the fractional linear polarization and the orientation of the linear polarization vectors around 
Stars~A and B are consistent with the measurements at larger scale (Yao et al. \cite{yao97}; Curran \& Chrysostomou \cite{cur07}),
and they measured a polarization $\rm{PA}$ of Star~A of $\rm{PA}^{\rm{Star~A}}_{P_{\rm{l}}}=-68^{\circ}\pm2$\d.
\begin{figure*}[th!]
\centering
\includegraphics[width = 6 cm]{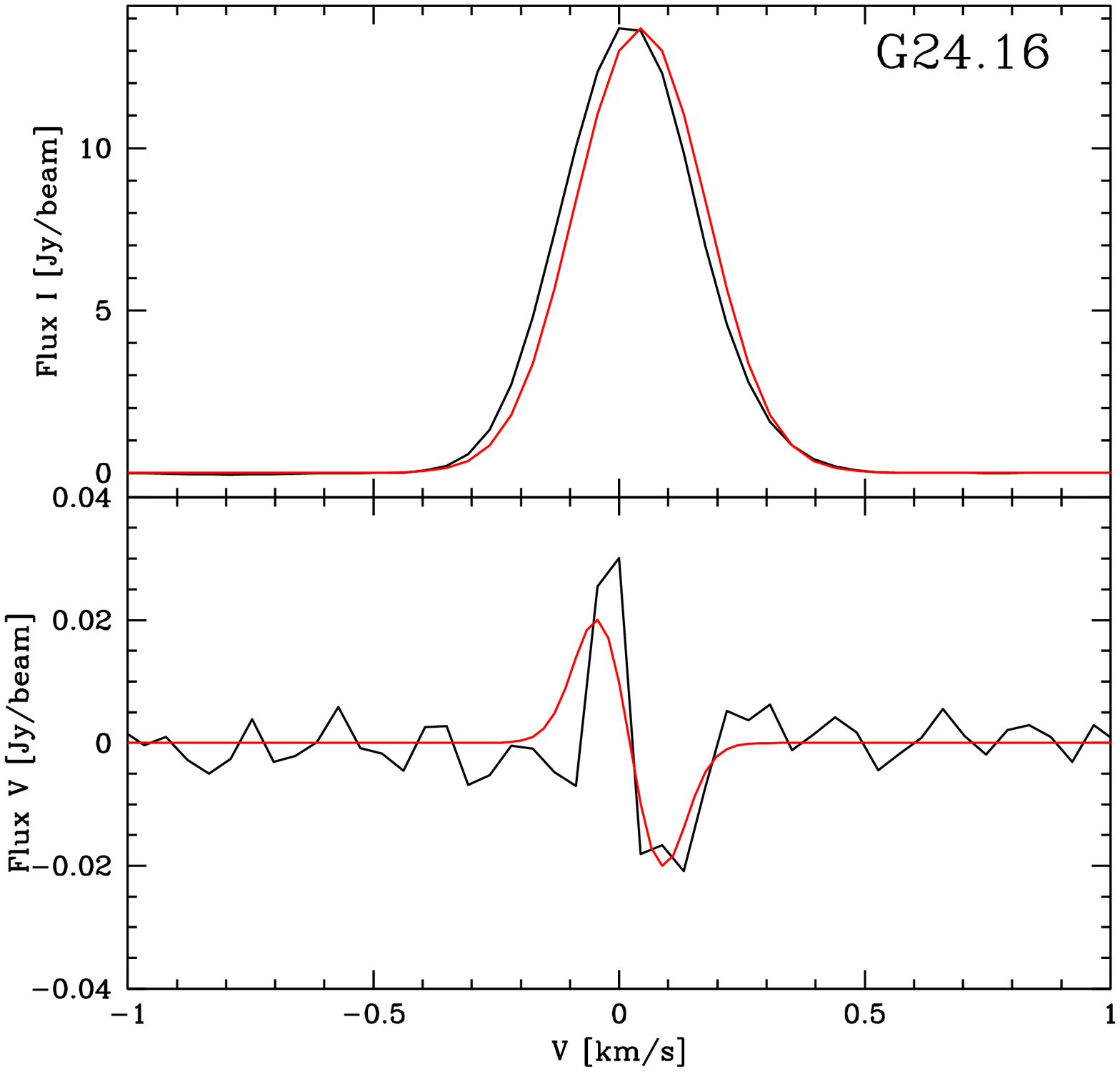}
\includegraphics[width = 6 cm]{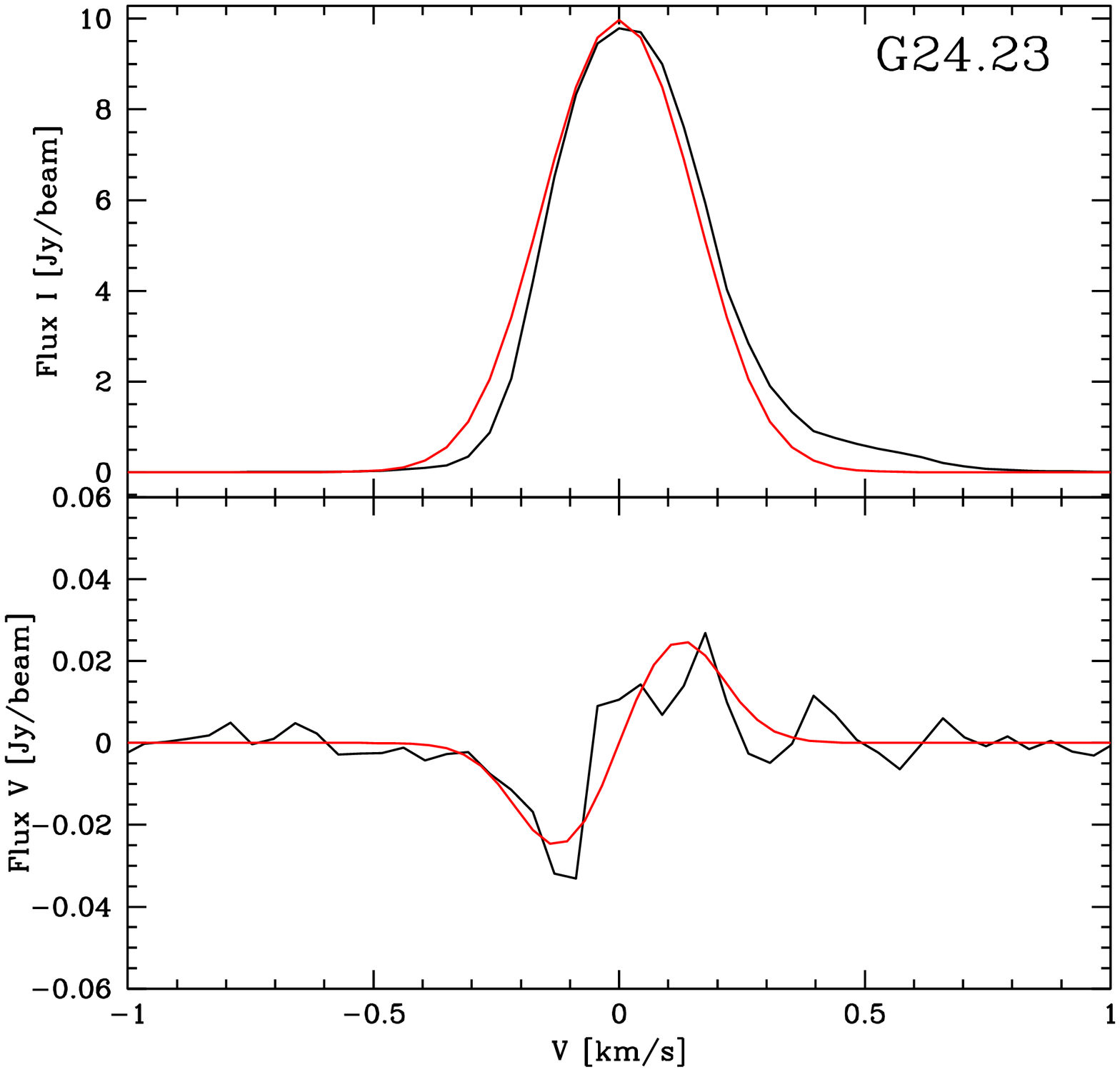}
\includegraphics[width = 6 cm]{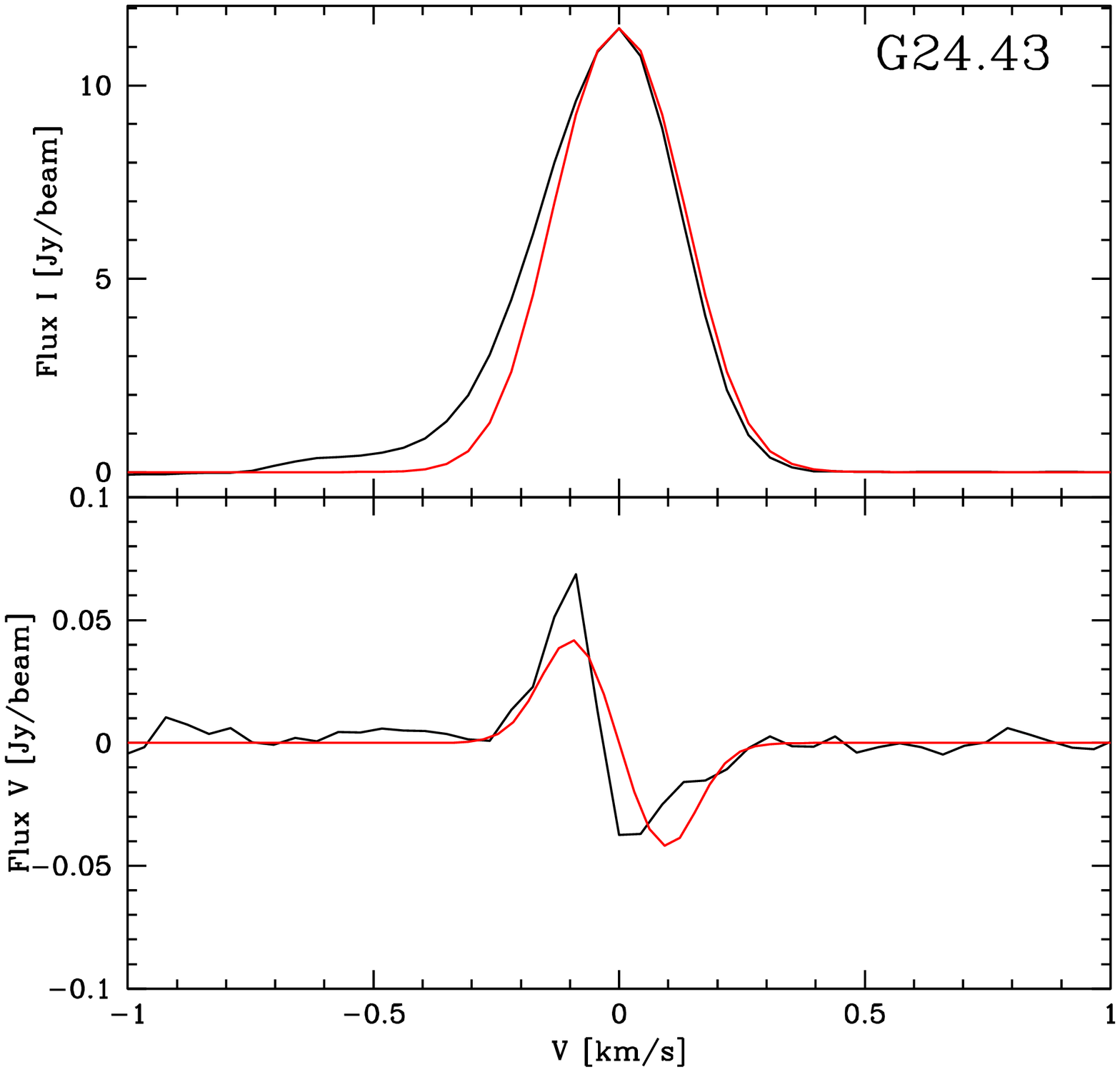}
\includegraphics[width = 6 cm]{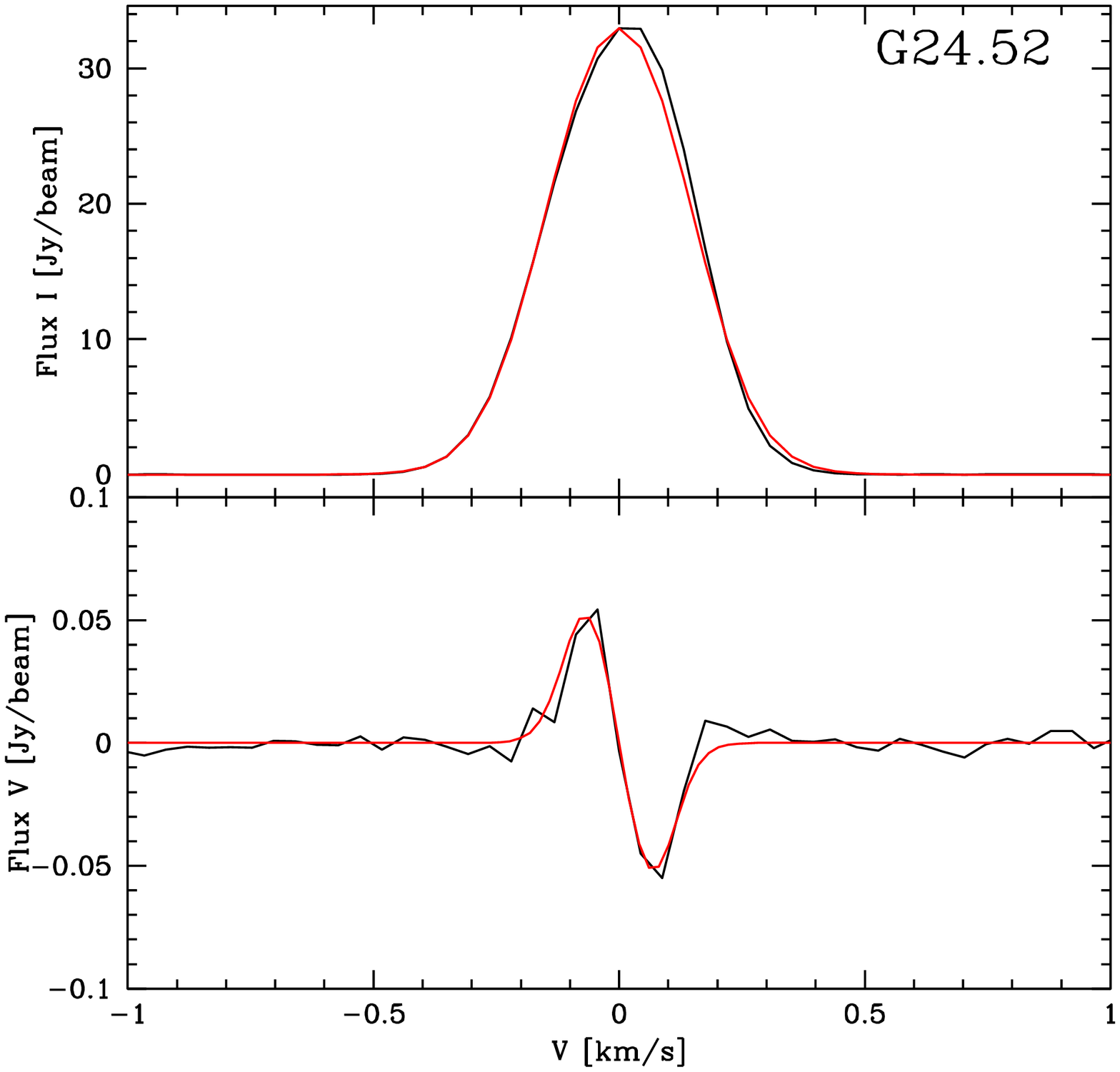}
\includegraphics[width = 6 cm]{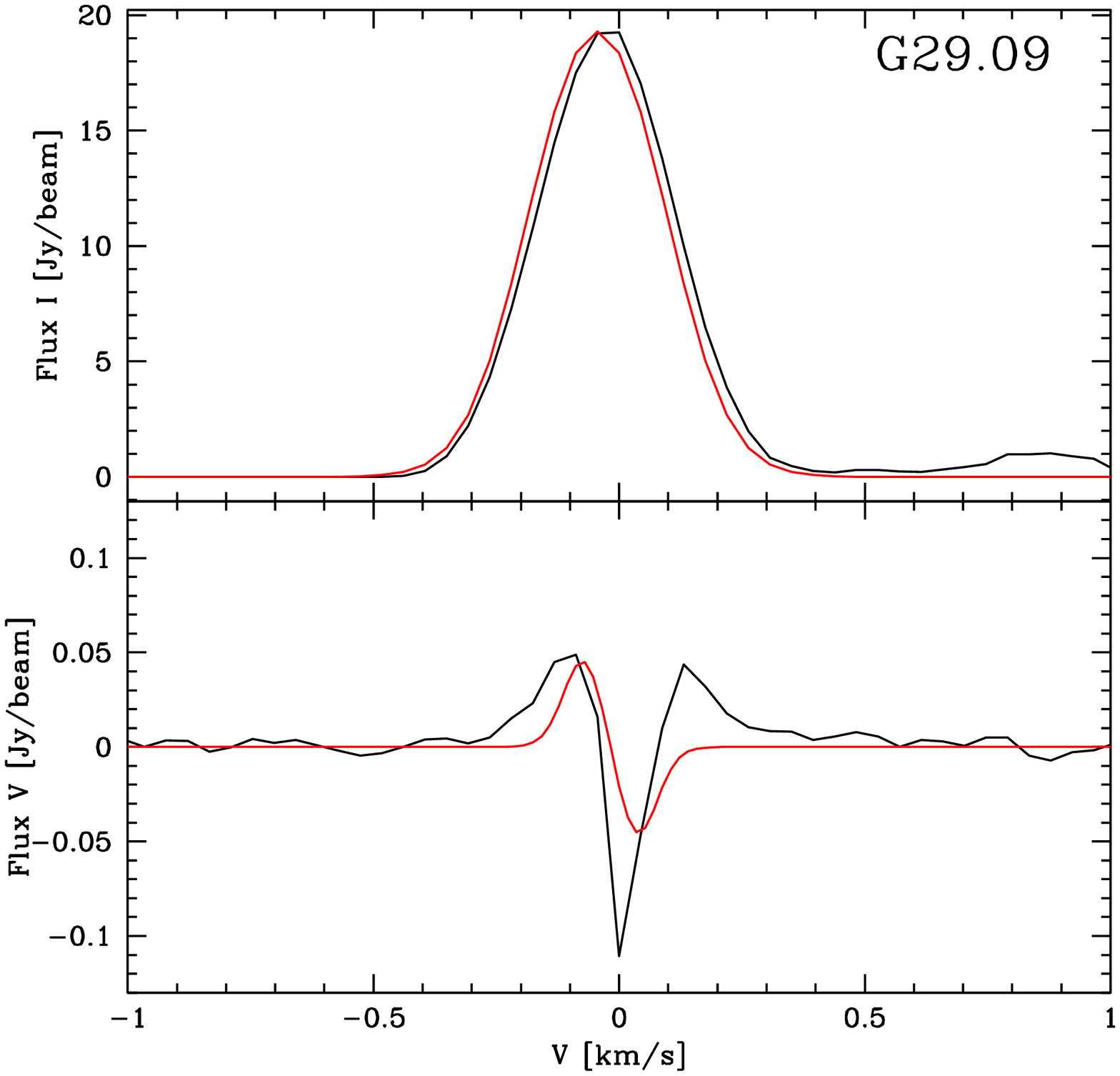}
\includegraphics[width = 6 cm]{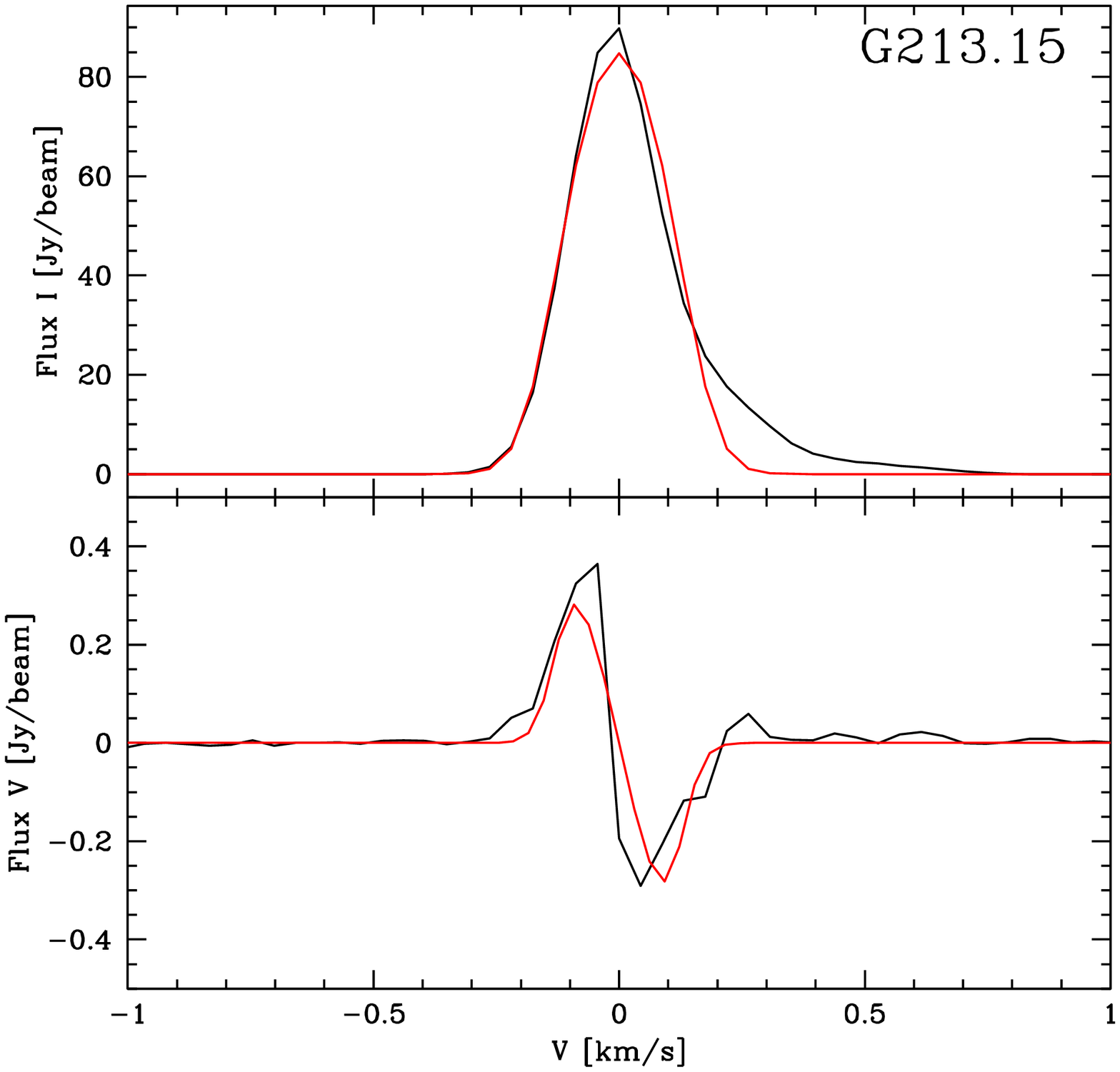}
\caption{Total intensity (\textit{I, upper panel}) and circularly polarized intensity (\textit{V, lower panel}) spectra for the \meth 
~maser features G24.16, G24.23, G24.43, G24.52, G29.09, and G213.15 (see Tables~\ref{G24_tab}, ~\ref{G29_tab}, and 
\ref{G213_tab}). 
The thick red lines are the best-fit models of \textit{I} and \textit{V} emission obtained using the adapted FRTM code (see 
Sect.~\ref{obssect}). The maser features were centered on zero velocity.}
\label{Vfit}
\end{figure*}
\section{Observations and analysis}
\label{obssect}
The first seven massive SFRs were observed at 6.7 GHz in full polarization spectral mode with eight of the EVN antennas 
(Effelsberg, Jodrell, Onsala, Medicina, Noto, Torun, Westerbork, and Yebes-40\,m) between November 2012 and June 2013, for a total 
observation time of 49~h. The bandwidth was 2~MHz, providing a velocity range of $\sim100$~\kms. The data were correlated with the 
EVN software correlator (SFXC; Keimpema et al. \cite{kei14}) at the Joint Institute for VLBI in Europe (JIVE) using 2048 channels and 
generating all four polarization combinations (RR, LL, RL, LR) with a spectral resolution of $\sim$1~kHz ($\sim$0.05~\kms). All the 
observational details are reported in Table~\ref{Obs}. We report
in Cols.~1 to 3 the target source, the 
program code, and the date of the observations; in Cols.~4 and 5 we list the polarization calibrators with their
polarization angles. Columns~6 to~8 list the restoring beam sizes, 
corresponding position angles, and the thermal noise. In Col.~9 we also show the self-noise
in the maser emission channels (see below for more details). Finally, Cols.~10 to 13 report the estimated absolute position of the 
reference maser and the FRMAP uncertainties (see below for more details).\\
\indent The data were edited and calibrated using AIPS. The bandpass, delay, phase, and polarization calibration were 
performed on the calibrators listed in Table~\ref{Obs}. Fringe-fitting and self-calibration were performed on the brightest maser 
feature of each star-forming region.  The \textit{I}, \textit{Q}, \textit{U}, and \textit{V} cubes were imaged using the AIPS task 
IMAGR. The \textit{Q} and \textit{U} cubes were combined to produce cubes of polarized intensity ($POLI=\sqrt{Q^{2}+U^{2}}$) 
and polarization angle ($POLA=1/2\times~atan(U/Q)$). We calibrated the linear polarization angles by comparing 
the linear polarization angles of the polarization calibrators measured by us with the angles obtained by calibrating the POLCAL 
observations made by NRAO\footnote{http://www.aoc.nrao.edu/$\sim$smyers/evlapolcal/polcal\_master.html}. The 
NRAO POLCAL observing program was temporarily interrupted because of the JVLA commissioning. The last POLCAL 
observations were made in May/June 2012, 
therefore we were able to calibrate the polarization angles of the sources observed in 2012 by using the results from the last 
observing run. The calibrator observed in 2013 was J2202+4216,
which  shows a constant polarization angle between 
2005\footnote{http://www.aoc.nrao.edu/$\sim$smyers/calibration/} and 2012 of -31\d$\pm4$\d. To calibrate the polarization 
angles of the maser sources observed in 2013, we therefore assumed that the 
polarization angle of J2202+4216 has not changed significantly. We were thus able to estimate the polarization angles with a systemic 
error of no more than $\sim$~5\d ~(see Col.~5 of Table~\ref{Obs}). The formal errors on $POLA$ are due to thermal noise. This error 
is given by $\sigma_{POLA}=0.5 ~(\sigma_{P}/POLI) \times (180^{\circ}/\pi)$ (Wardle \& Kronberg \cite{war74}), where 
$\sigma_{P}$ is the rms error of POLI.\\
\indent Because the observations were not performed in phase-referencing mode, we estimated the absolute position of the 
brightest maser feature of each source through fringe rate mapping by using the AIPS task FRMAP. The results and the formal errors of 
FRMAP are reported in Cols.~10 to 13 of Table~\ref{Obs}. The absolute positional uncertainties are dominated 
by the phase fluctuations that we estimate to be on the order of no more than a few mas from our experience with other experiments 
and varying the task parameters.\\
\indent We analyzed the polarimetric data following the procedure reported in Papers~I and~II. First, 
we identified the \meth ~maser features by using the process described in Surcis et al. (\cite{sur11b}), and then we determined the 
mean linear polarization fraction ($P_{\rm{l}}$) and the mean linear polarization angle ($\chi$) across the spectrum of each \meth 
~maser feature. Second, we made use of the adapted full radiative transfer method (FRTM) code for 6.7 GHz \meth ~masers (Vlemmings
et al. \cite{vle10}, Surcis et al. \cite{sur11a}, Paper~II) to model the total intensity and the linearly polarized spectrum 
of every maser feature for which we were able to detect linearly polarized emission. The output of this code provides estimates of the 
emerging brightness temperature (\tbo) and of the intrinsic thermal line width (\dvi). Following Surcis et al. (\cite{sur11a}), we 
restricted our analysis to values of \dvi ~from 0.5~\kms ~to 1.95~\kms. From \tbo ~and $P_{\rm{l}}$, we then determined the angle 
between the propagation direction of the maser radiation and the magnetic field ($\theta$). If 
$\theta>\theta_{\rm{crit}}=55$\d, where 
$\theta_{\rm{crit}}$ is the Van Vleck angle, the magnetic field appears to be perpendicular to the linear polarization vectors; 
otherwise, it is parallel (Goldreich et al. \cite{gol73}). To better determine the orientation of the magnetic field with respect to 
the linear polarization vectors, we followed the method introduced in Paper~II that takes into consideration the 
errors associated with $\theta$, that is, $\varepsilon^{\rm{\pm}}$. According to this, the magnetic field is most likely  perpendicular to the linear 
polarization vectors if $|\theta^{\rm{+}}-55$\d$|>|\theta^{\rm{-}}-55$\d$|$, where $\theta^{\rm{\pm}}=\theta\pm\varepsilon^{\rm{\pm}}$;
otherwise, the magnetic field is assumed to be parallel. Of course, if $\theta^{\rm{-}}$ and $\theta^{\rm{+}}$ are either larger or 
smaller than 55\d ~, the magnetic field is perpendicular or parallel to the linear polarization vectors, respectively.\\
\indent Note that if $T_{\rm{b}}\Delta\Omega>2.6\times10^9~\rm{K~sr}$ the 6.7 GHz 
\meth ~masers can be considered partially saturated and their \dvi ~and \tbo ~values are overestimated and 
underestimated, respectively (Surcis et al. \cite{sur11a}). However, we are confident that the orientation of their linear polarization vectors 
is not affected by their saturation state (Paper~I), and consequently, they can be used for 
determining the orientation of the magnetic field in the region.\\
\indent Finally, to measure the Zeeman splitting (\dvz), we included the best estimates of 
\tbo ~and \dvi ~in the \code ~to produce 
the $I$ and $V$ models used for fitting the total intensity and circularly polarized spectra of the corresponding \meth ~maser feature
(Fig.~\ref{Vfit}). Because the circularly polarized emission of \meth ~masers is usually very weak ($<1\%$), we must take into 
consideration the self noise\footnote{The self-noise is high when the power contributed by the astronomical maser is a significant
portion of the total received power (Sault \cite{sau12}).} ($\sigma_{\rm{s.-n.}}$) produced by the masers (Col.~9 of Table~\ref{Obs}; 
e.g., Sault \cite{sau12}) when we measure the Zeeman splitting. Therefore, we consider real a detection of circularly polarized emission only 
when the detected $V$ peak flux of a maser feature is both five times higher than the rms and three times larger than 
$\sigma_{\rm{s.-n.}}$. We know
from the Zeeman effect theory that \dvz ~is related to the magnetic field strength along the line 
of sight ($B_{||}$) through $\Delta V_{\rm{Z}}=\alpha_{\rm{Z}}\cdot B_{||}$. However,
 the Land\'{e} g-factors for the \meth ~molecule (including the 6.7 GHz maser transition) on which $\alpha_{\rm{Z}}$ depends are 
still unknown, and consequently, the magnetic field strength cannot yet be derived from our Zeeman-splitting measurements (e.g., 
Vlemmings et al. \cite{vle11}).
\section{Results}
\label{res}
\begin{figure*}[t!]
\centering
\includegraphics[width = 9 cm]{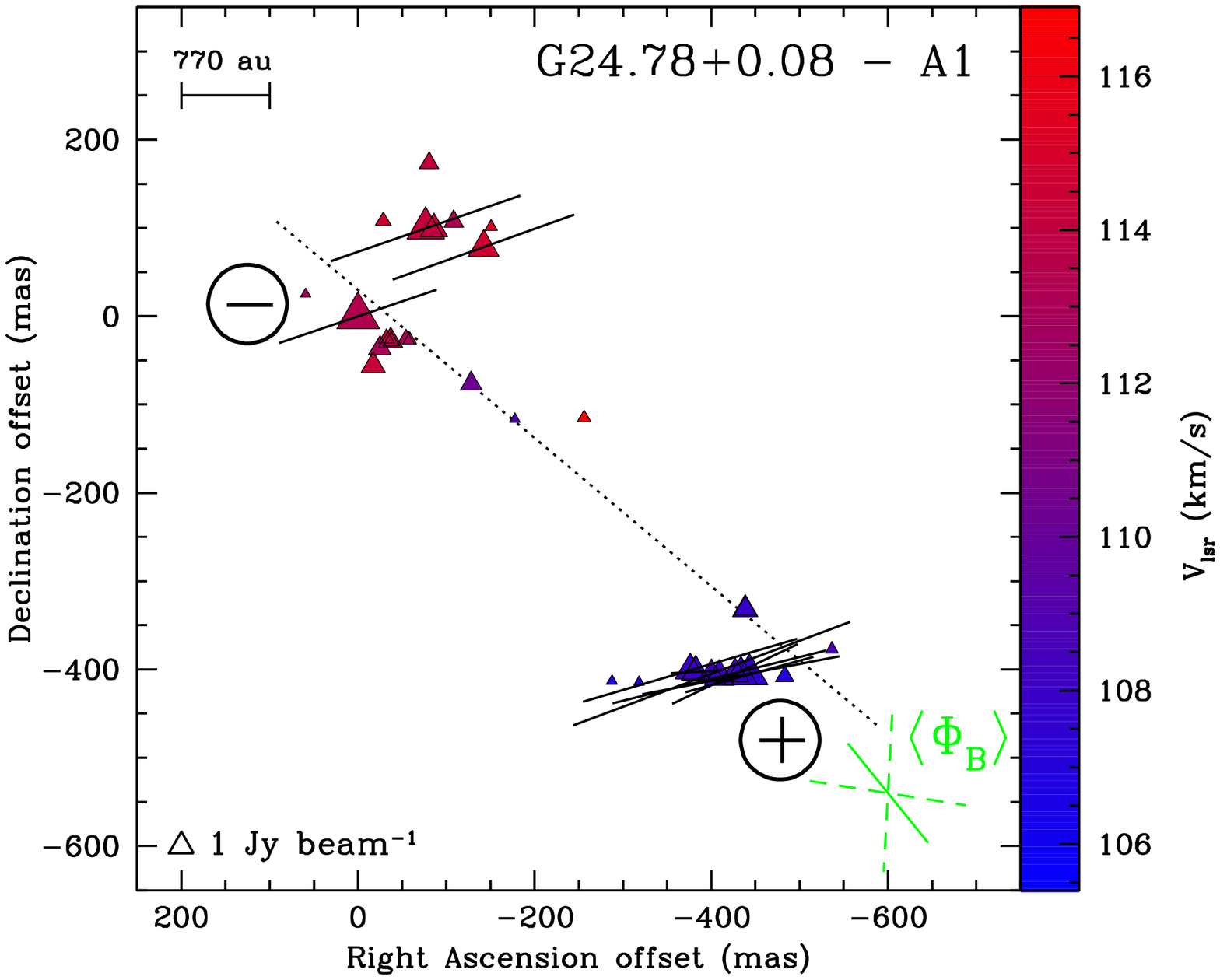}
\includegraphics[width = 9 cm]{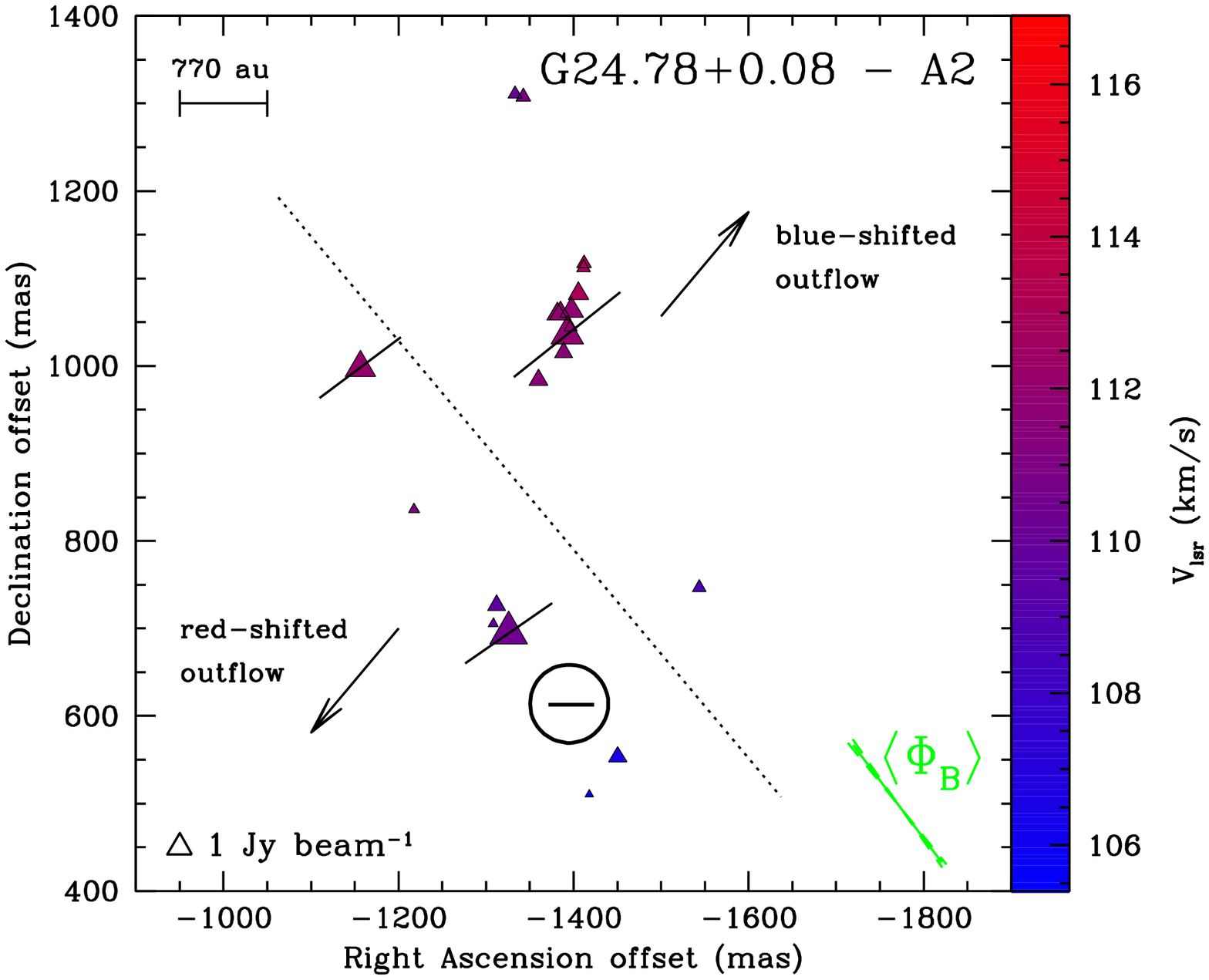}
\caption{View of the \meth ~maser features detected around the mm subcores A1 (left panel) and A2 (right panel) of G24.78+0.08.
The reference position is $\alpha_{2000}=+18^{\rm{h}}36^{\rm{m}}12^{\rm{s}}\!.563$ and 
$\delta_{2000}=-07^{\circ}12'10''\!\!.787$ (see Sect.~\ref{obssect}).
Triangle symbols identify \meth ~maser features scaled logarithmically according to their peak flux density (Table~\ref{G24_tab}). 
Maser LSR radial velocities are indicated by color (the assumed velocity of the region is $V_{\rm{lsr}}^{\rm{^{12}CO(1-0)}}=+111$~\kms,
Furuya et al. \cite{fur02}). A 1~\jyb ~symbol is plotted for illustration in both panels. The linear polarization vectors, scaled 
logarithmically according to polarization fraction $P_{\rm{l}}$, are overplotted. In the bottom right corner of both panels 
the 
corresponding error-weighted orientation of the magnetic field ($\langle\Phi_{\rm{B}}\rangle$, see Sect.\ref{Borient}) is also reported, 
the two dashed segments indicate the uncertainty. The two arrows in the right panel indicate the direction, 
and not the actual position, of the 
red- and blueshifted lobes of the $\rm{^{12}CO(1-0)}$ outflow associated with G24.78+0.08-A2 ($\rm{PA_{outflow}^{^{12}CO}}=-40$\d; 
Beltr\'{a}n et al. \cite{bel11}). The dotted lines indicate the direction of the CH$_3$CN toroids (Beltr\'{a}n et al. 
\cite{bel11}). The circled plus and minus symbols indicate where the magnetic field points away from and 
where toward the observer, respectively.}
\label{G24_cp}
\end{figure*}

In Tables~\ref{G24_tab}--\ref{G213_tab} we list all the 6.7 GHz \meth ~maser features detected toward the seven massive SFRs 
observed with the EVN. The description of the maser distribution and the polarization results are reported for each source 
separately in Sects.~\ref{G24_sec}--\ref{G213_sec}. In Figs. \ref{G24_cp}--\ref{G213_cp} we show the measured linear 
polarization vectors as black segments and the inferred orientation of the magnetic field, which is either parallel or perpendicular to 
the linear polarization vectors (see Sect.~\ref{obssect}), in green in the bottom right corner of each panel.
\subsection{\object{G24.78+0.08}}
\label{G24_sec}
We detected 53 \meth ~maser features, named G24.01--G24.53 in Table~\ref{G24_tab}, 33 toward core A1 and 20 
toward A2.
In Fig.~\ref{G24_cp} we show all the maser features associated with A1 in the left panel and those associated with
A2 in the right panel.
The maser distributions around the two cores are identical to those observed previously by Moscadelli et al. (\cite{mos07}), even 
though we detected about 40 maser features more. The peak flux density range (Col.~5 in Table~\ref{G24_tab}) and the local 
standard of rest velocity ($V_{\rm{lsr}}$; Col.~6 in Table~\ref{G24_tab}) range are similar to previous measurements.\\
\indent We detected linearly polarized emission from ten \meth ~maser features around A1 ($P_{\rm{l}}^{\rm{A1}}=0.8\%-3.5\%$) and 
from three maser features around A2 ($P_{\rm{l}}^{\rm{A2}}=1.0\%-1.3\%$). The adapted \code ~was able to fit all of them but G24.38. The 
outputs of the code are reported in Cols.~10, 11, and 14 of Table~\ref{G24_tab}. The twelve maser features for which we estimated \tbo 
~are unsaturated. Indeed, $T_{\rm{b}}\Delta\Omega<2.6\times10^9$~K~sr (or in logarithmic value $<9.4$~log~K~sr). For the maser features 
G24.23 and G24.52, both associated with A1, we have that $|\theta^{\rm{+}}-55$\d$|<|\theta^{\rm{-}}-55$\d$|$, that is, the magnetic field is 
assumed to be parallel to their linear polarization vectors as described in Sect.~\ref{obssect}. We also measured \dvz ~for five \meth 
~maser features (Col.~13 of Table~\ref{G24_tab}), only one of
which is associated with A2. The circular polarization fraction 
($P_{\rm{V}}$) ranges from 0.3\% to 0.7\% and the Zeeman splitting is $-9.7$~\ms$\leq\Delta V_{\rm{Z}}^{\rm{A1}}\leq+7.8$~\ms ~around
A1 and $\Delta V_{\rm{Z}}^{\rm{A2}}=(-4.0\pm0.8)$~\ms ~around A2 (see Fig.~\ref{Vfit}).
\subsection{\object{G25.65+1.05}}
\label{G25_sec}
\begin{figure}[t!]
\centering
\includegraphics[width = 9 cm]{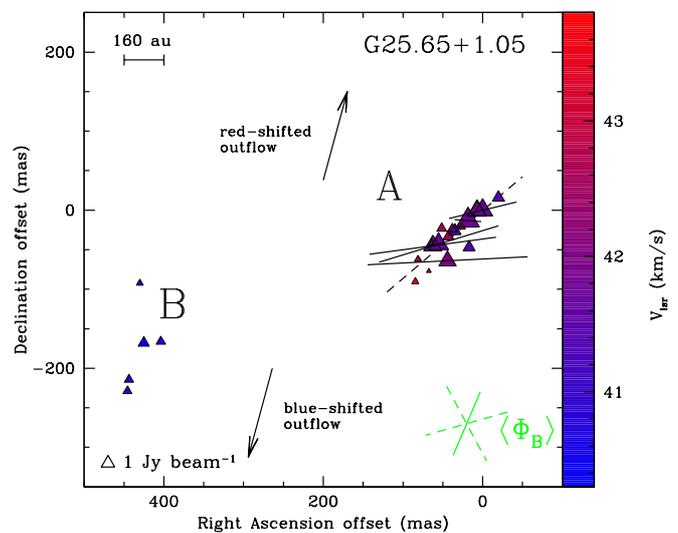}
\caption{View of the \meth ~maser features detected around G25.65+1.05 (Table~\ref{G25_tab}). Same symbols as in Fig.~\ref{G24_cp}.
The assumed velocity of the YSO is $V_{\rm{lsr}}^{\rm{N_{2}H^+-C_{2}H}}=+42.41$~\kms ~(S\'{a}nchez-Monge et al. \cite{san13}). The two 
arrows indicate the direction, and not the actual position, of the red- and blueshifted lobe of the bipolar outflow 
($\rm{PA_{outflow}^{\rm{SiO}}}=-15$\d; 
S\'{a}nchez-Monge et al. \cite{san13}). The dashed line is the best linear fit of the \meth ~maser features of group A 
($\rm{PA_{CH_{3}OH}}=-49^{\circ}\pm7$\d).} 
\label{G25_cp}
\end{figure}
Imaging a $2''\times~2''$ field-of-view centered on G25.02, we were able to detect a total of 23 6.7-GHz \meth ~maser
features, named G25.01--G25.23 in Table~\ref{G25_tab}. The maser features can be divided into two groups (named here group~A and 
group~B) separated from each other by about 400~mas ($\sim$1300~au; see Fig.~\ref{G25_cp}). The two groups are located at the origin of 
the bipolar outflow. Comparing our detections with 
the four \meth ~maser spots detected by Walsh 
et al. (\cite{wal98}), which were linearly distributed southwards over $1''$, we note that only group~A can be associated with one 
of the previous maser spots, spot B (as named by Walsh et al. \cite{wal98}). The other three \meth ~maser 
spots were not detected by us, and group~B was not detected by Walsh et al. (\cite{wal98}). All the maser features of 
group~B are blueshifted with respect to the systemic velocity ($V_{\rm{lsr}}^{\rm{N_{2}H^+-C_{2}H}}=+42.41$~\kms; 
S\'{a}nchez-Monge et al. \cite{san13}). Even though the maser features of group~A show both a linear distribution 
($\rm{PA_{CH_{3}OH}}=-49^{\circ}\pm7$\d) and red- and blueshifted velocities, no clear velocity gradient is observed.\\
\indent Five \meth ~maser features of group~A show linearly polarized emission ($P_{\rm{l}}=0.3\%-1.3\%$), and according to the 
output of the adapted \code ~all of them are unsaturated (see Col.~11 of Table~\ref{G25_tab}). For the maser features G25.02 and G25.06 
we determined that $|\theta^{\rm{+}}-55$\d$|<|\theta^{\rm{-}}-55$\d$|$, that is, the magnetic field is assumed to be parallel to their 
linear polarization vectors. For the other maser features we found that $|\theta^{\rm{+}}-55$\d$|>|\theta^{\rm{-}}-55$\d$|$. 
We did not detect any circularly polarized maser emission toward the region ($P_{\rm{V}}<1.5\%$).
\subsection{\object{G29.86-0.04}}
\label{G29_sec}
\begin{figure}[t!]
\centering
\includegraphics[width = 9 cm]{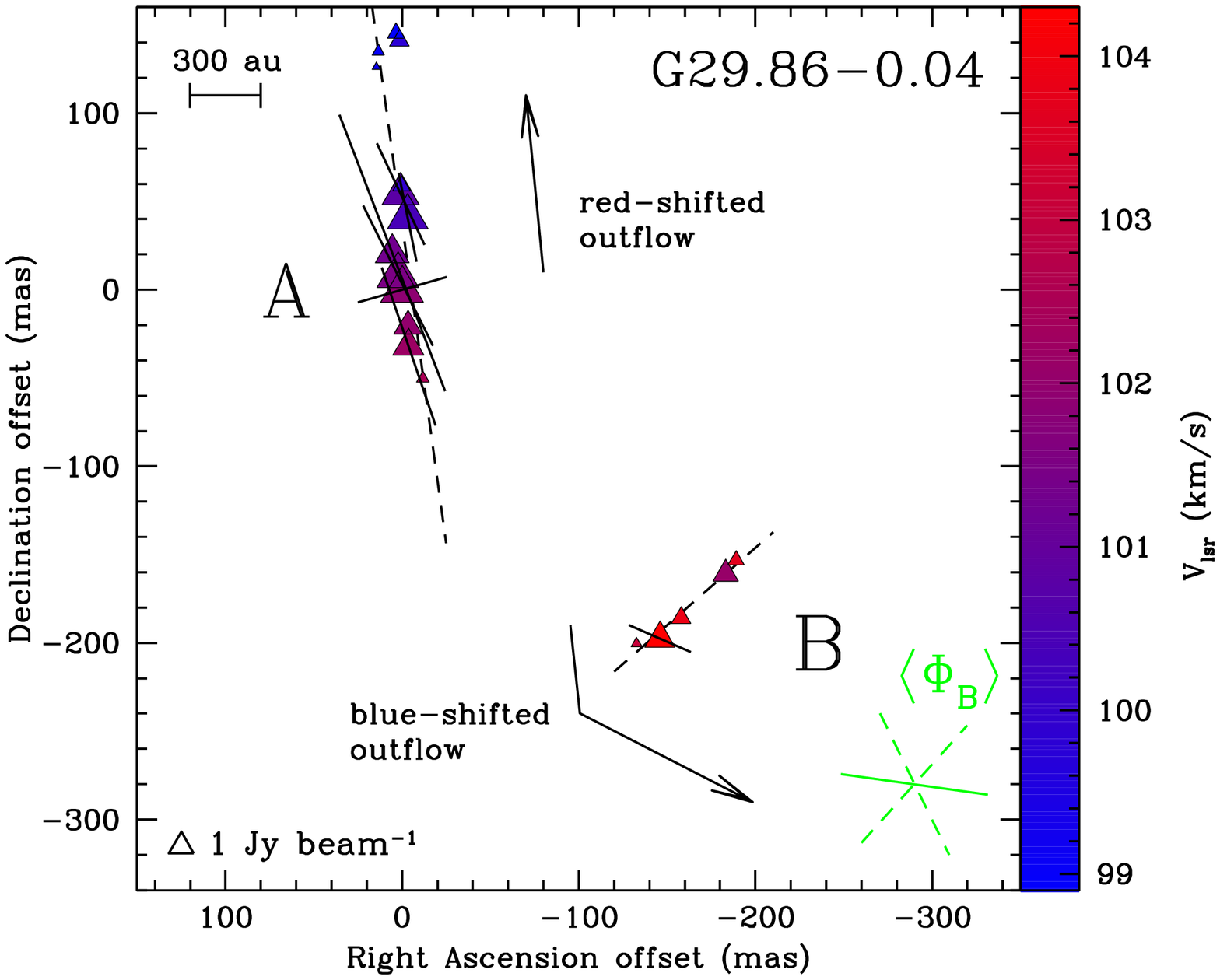}
\caption{View of the \meth ~maser features detected around G29.86-0.04 (Table~\ref{G29_tab}). Same symbols as in Fig.~\ref{G24_cp}.
The assumed velocity of the massive SFR region is $V_{\rm{lsr}}^{\rm{C^{18}O}}=+101.85$~\kms ~(de Villiers et al. \cite{dev14}). The 
two arrows indicate the direction, and not the actual position, of the red- and blueshifted lobe of the bipolar outflow 
($\rm{PA_{red-shifted}^{\rm{^{13}CO}}}\approx+6$\d ~and $\rm{PA_{blue-shifted}^{\rm{^{13}CO}}}\approx+60$\d;
de Villiers et al. \cite{dev14}). The dashed lines are the best linear fit to the positions of the \meth ~maser features 
of group~A and B ($\rm{PA_{CH_{3}OH}^{A}}=+8^{\circ}\pm7$\d ~and $\rm{PA_{CH_{3}OH}^{B}}=-49^{\circ}\pm5$\d).} 
\label{G29_cp}
\end{figure}
In Table~\ref{G29_tab} and Fig.~\ref{G29_cp} we report the 18 \meth ~maser features that we detected in the region. We
divided the maser features into two groups (A and B), and from a linear fit we find that the features of group~A are aligned with 
the redshifted lobe of the outflow ($\rm{PA_{CH_{3}OH}^{A}}=+8^{\circ}\pm7$\d and $\rm{PA_{red-shifted}^{\rm{^{13}CO}}}\approx+6$\d).
The five maser features of group~B are instead linearly distributed perpendicularly to the blueshifted lobe of the outflow
($\rm{PA_{CH_{3}OH}^{B}}=-49^{\circ}\pm5$\d and $\rm{PA_{blue-shifted}^{\rm{^{13}CO}}}\approx+63$\d), even though they are 
spatially associated with the redshifted lobe (see Fig.~B-1 of de Villiers et al. \cite{dev14}). The maser features of group~A
show a velocity gradient, from north (the most blueshifted velocity) to south (the most redshifted velocity), 
the range of which is consistent with the velocities of the quiescent emission of $\rm{^{13}CO}$ 
($+96.5$~\kms$<V_{\rm{quiescent}}^{\rm{^{13}CO}}<+104$~\kms; de Villiers et al. \cite{dev14}). 
 The velocity range of group~B is also consistent with $V_{\rm{quiescent}}^{\rm{^{13}CO}}$.\\
\indent Almost 40\% of the \meth ~maser features show linearly polarized emission ($P_{\rm{l}}=1.2\%-17\%$) and only 
the highest linearly polarized feature (i.e., G29.17 in Table~\ref{G29_tab}) appears to have a high saturation degree 
($T_{\rm{b}}\Delta\Omega=6.3\times10^{10}$~K~sr). From our analysis of the estimated $\theta$ values we determined that the magnetic
field is perpendicular to all the maser features but G29.14, for which $\theta=67^{\circ~+10^{\circ}}_{-46^{\circ}}$. We measured a 
Zeeman splitting of $\Delta V_{\rm{Z}}=-6.6\pm1.1$~\ms ~toward G29.09 ($P_{\rm{V}}=0.5\%$; see Fig.~\ref{Vfit}).
\subsection{\object{G35.03+0.35}}
\label{G35_sec}
\begin{figure}[t!]
\centering
\includegraphics[width = 9 cm]{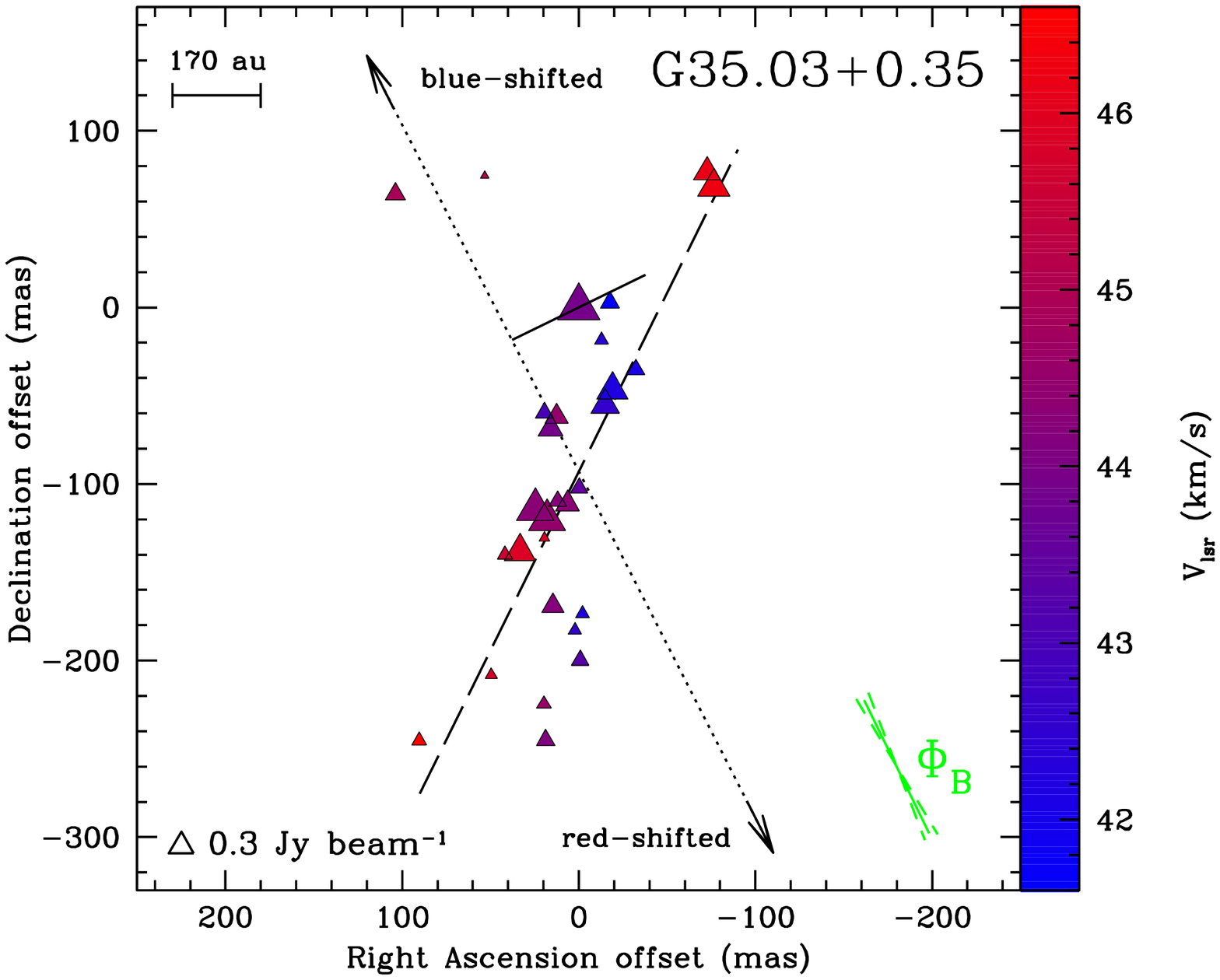}
\caption{View of the \meth ~maser features detected around G35.03+0.35 (Table~\ref{G35_tab}). Same symbols as in Fig.~\ref{G24_cp}.
The assumed velocity of the massive SFR region is $V_{\rm{lsr}}=+51.5$~\kms ~(Paron et al. \cite{par12}). The dotted line indicates 
the direction, and not the actual position, of the bipolar 4.5~$\mu$m emission ($\rm{PA^{4.5\,\mu m}}\approx+27^{\circ}$; 
Cyganowski et al. \cite{cyg09}) and the two arrows indicate the direction of the red- and blueshifted lobes of the $^{12}$CO-outflow 
(Paron et al. \cite{par12}).
The dashed line is the best linear fit of the \meth ~maser features ($\rm{PA_{CH_{3}OH}}=-26^{\circ}\pm19$\d).} 
\label{G35_cp}
\end{figure}
Across a bandwidth that covers a range of velocities between $-6$~\kms ~and +94~\kms ~, we detected 29 6.7-GHz \meth ~maser features 
with velocities +41.5~\kms$<V_{\rm{lsr}}<$+46.7~\kms (see Table~\ref{G35_tab}). No redshifted features were detected, as previously 
reported (Szymczak et al. \cite{szy00}; Pandian et al. \cite{pan11}). The maser features (Fig.~\ref{G35_cp}) are distributed from 
southeast to northwest ($\rm{PA_{CH_{3}OH}}=-26^{\circ}\pm19$\d) almost perpendicular to the 4.5~$\mu$m emission.
In Fig.~\ref{G35_cp} we have drawn the two arrows assuming that the bipolar 4.5~$\mu$m emission also traces the
$^{12}$CO large-scale outflow (Cyganowski et al. \cite{cyg09}, Paron et al. \cite{par12}).\\
\indent Because of the weak 6.7 GHz \meth ~maser features, we were able to measure linear polarization only toward the brightest 
maser feature G35.19 ($P_{\rm{l}}=0.9\%$), which appears to be unsaturated. The corresponding $\theta$ angle was 
$+90^{\circ}\pm24$\d. No circularly polarized emission was detected ($P_{\rm{V}}<0.8\%$).
\subsection{\object{G37.43+1.51}}
\label{G37_sec}
\begin{figure}[t!]
\centering
\includegraphics[width = 9 cm]{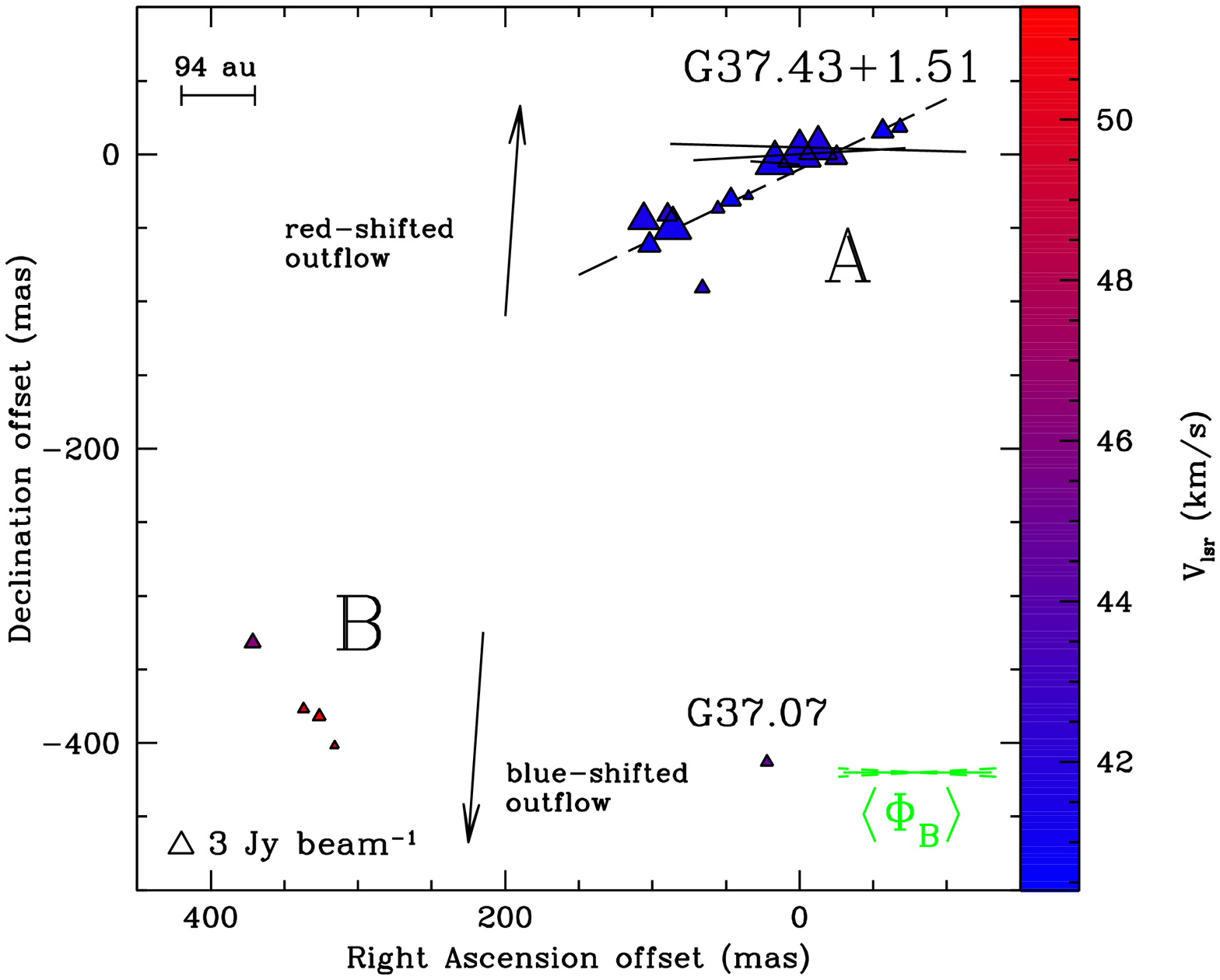}
\caption{View of the \meth ~maser features detected around G37.43+1.51 (Table~\ref{G37_tab}). Same symbols as in Fig.~\ref{G24_cp}.
The assumed velocity of the massive SFR region is $V_{\rm{lsr}}^{\rm{C^{18}O}}=+44.1$~\kms ~(L\'{o}pez-Sepulcre et al. \cite{sep10}). 
The two arrows indicate the direction, and not the actual position, of the red- and blueshifted lobes of the 
C$^{18}$O-outflow 
($\rm{PA_{outflow}^{\rm{C^{18}O}}}=-4$\d; L\'{o}pez-Sepulcre et al. \cite{sep10}). The dashed line is the best linear fit of the 
\meth ~maser features ($\rm{PA_{CH_{3}OH}}=-64^{\circ}\pm5$\d).} 
\label{G37_cp}
\end{figure}
We detected two groups of \meth ~maser features, named group~A and group~B in Fig.~\ref{G37_cp}, separated by 300~mas ($\sim$550~au). 
Group~A is composed of 14 maser features distributed linearly with $\rm{PA_{CH_{3}OH}}=-64^{\circ}\pm5$\d ~with no clear velocity 
gradient, as already reported by Fujisawa et al. (\cite{fuj14}). The velocities of group~A are consistent with the velocity 
range of
the blueshifted lobe of the C$^{18}$O-outflow (L\'{o}pez-Sepulcre et al. \cite{sep10}). Maser features of group~B were not detected 
before. This group, which is located southeast w.r.t. group~A, show a velocity range of between 46~\kms ~and 52~\kms. Furthermore, an 
isolated maser feature (G37.07; see Table~\ref{G37_tab}) that cannot be associated with either of the two groups is 
located at 400~mas ($\sim$750~au) south and 300~mas ($\sim$560~au) west from groups~A and~B, respectively. \\
\indent We detected linearly polarized emission from three \meth ~maser features of group~A ($P_{\rm{l}}=0.7\%-1.5\%$), all of which 
have an estimated \tbo ~lower than the saturation threshold. The \code ~estimated that the magnetic 
field is parallel to all the linear polarization vectors of these features, indeed $|\theta^{\rm{+}}-55$\d$|<|\theta^{\rm{-}}-55$\d$|$ 
(see Col.~14 of Table~\ref{G37_tab}). No circular polarization was measured ($P_{\rm{V}}<0.2\%$).
\subsection{\object{G174.20-0.08}}
\label{G174_sec}
\begin{figure}[t!]
\centering
\includegraphics[width = 9 cm]{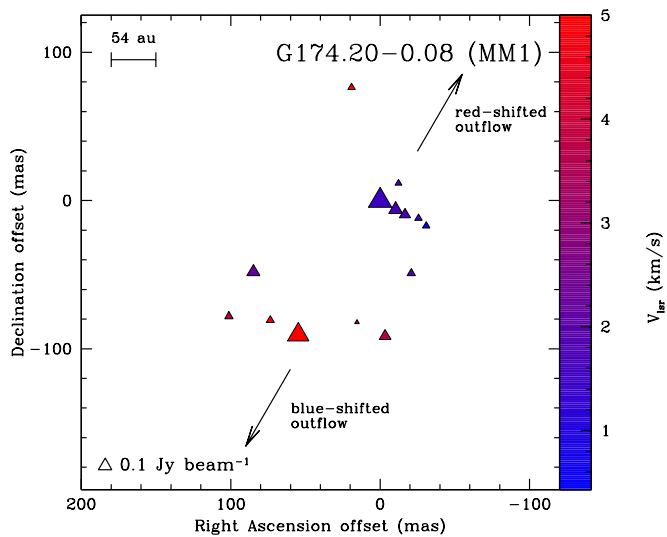}
\caption{View of the \meth ~maser features detected around the millimeter core MM1 of G174.20-0.08 (Table~\ref{G174_tab}). Same 
symbols as in Fig.~\ref{G24_cp}. The assumed velocity of the massive SFR region is $V_{\rm{lsr}}^{\rm{CH_{3}CN}}=-1.0$~\kms 
~(Zhang et al. \cite{zha07}). 
The two arrows indicate the direction, and not the actual position, of the red- and blueshifted lobes of the 
$^{12}$CO-outflow ($\rm{PA_{outflow}^{\rm{H_{2}O}}}=-40$\d; Goddi et al. \cite{god11}).} 
\label{G174_cp}
\end{figure}
The 14 6.7-GHz \meth ~maser features detected toward AFGL\,5142 are shown in Fig.~\ref{G174_cp}. No \meth ~maser emission with a peak 
flux density $>0.9$~\jyb ~was detected. Both the maser distribution and the velocity range of the masers agree with previous 
observations (e.g., Goddi et al. \cite{god11}). We were not able to detect at 5$\sigma$ either linearly polarized 
($P_{\rm{l}}<0.02\%$) or circularly polarized maser emissions ($P_{\rm{V}}<0.02\%$).
\subsection{\object{G213.70-12.6}}
\label{G213_sec}
\begin{figure}[t!]
\centering
\includegraphics[width = 9 cm]{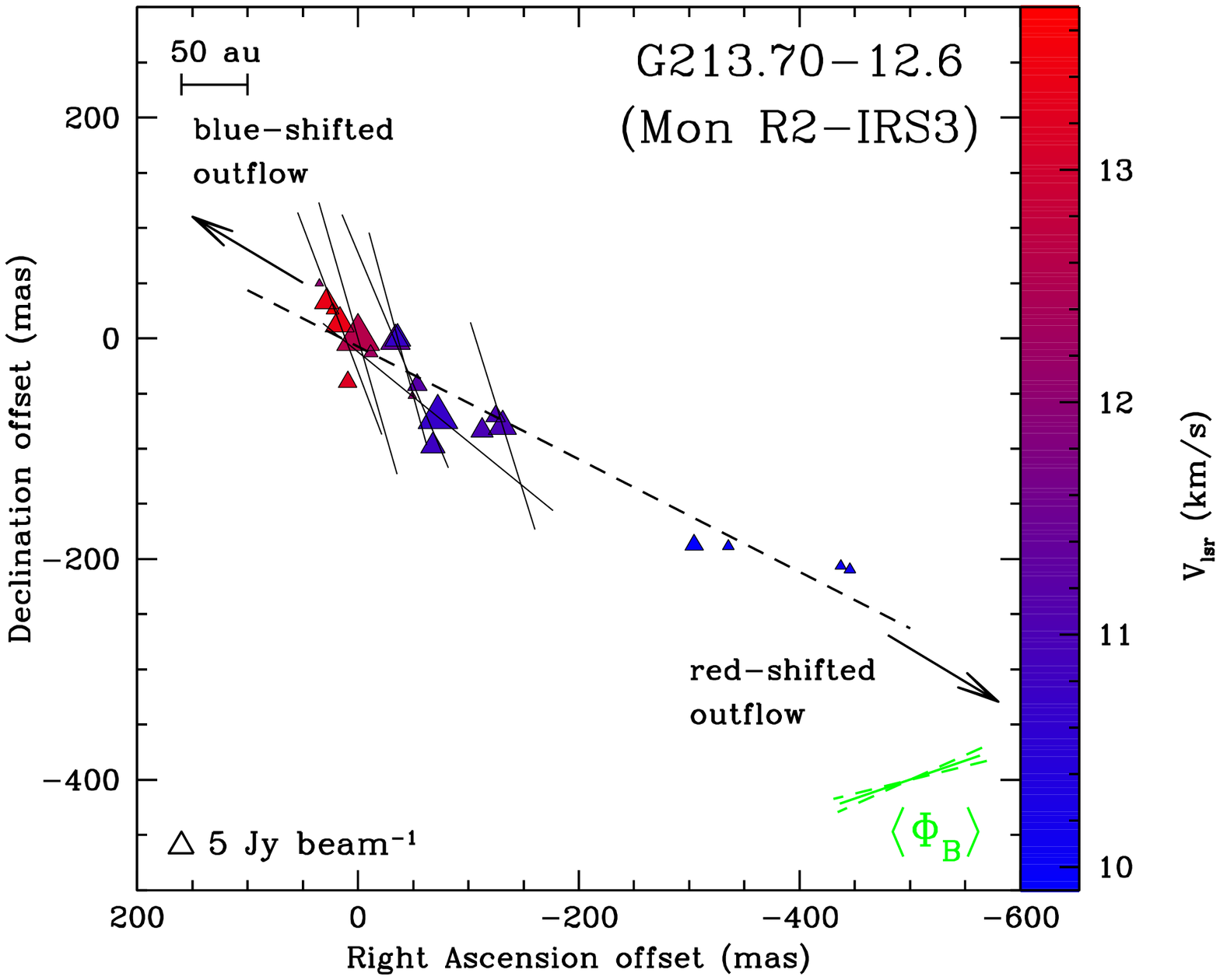}
\caption{View of the \meth ~maser features detected toward Star~A in the massive SFR G213.70-12.6 (Mon~R2-IRS3; Table~\ref{G213_tab}). 
Same symbols as in Fig.~\ref{G24_cp}. The assumed velocity of the massive SFR region is $V_{\rm{lsr}}^{\rm{^{13}CO(2-1)}}=+10.5$~\kms 
~(Dierickx et al. \cite{die15}). The arrows indicate the direction, and not the actual position,  of the red- and blueshifted 
$^{13}$CO(2-1)-outflow powered by IRS\,3 ($\rm{PA_{outflow}^{\rm{^{13}CO(2-1)}}}=+53$\d; Dierickx et al. \cite{die15}). The dashed line 
is the best linear fit of the \meth ~maser features ($\rm{PA_{CH_{3}OH}}=+63^{\circ}\pm2$\d).
} 
\label{G213_cp}
\end{figure}
We detected 20 \meth ~maser features that are linearly distributed from northeast to southwest with 
$\rm{PA_{CH_{3}OH}}=+63^{\circ}\pm2$\d (see Fig.~\ref{G213_cp}). Because the most western maser features (G213.01--G213.04 in 
Table~\ref{G213_tab}) were previously undetected (Minier et al. \cite{min00}), the linear distribution of the maser features
is now more extended (545~mas; $\sim$450~au).\\
\indent Six of the \meth ~maser features showed linearly polarized emission ($P_{\rm{l}}=3.0\%-5.0\%$), and according to the estimated
\tbo ~three of them are unsaturated (G213.08, G213.12, and G213.13). The \code ~estimated $\theta$ angles greater than 55\d ~, indicating
that the magnetic field is perpendicular to all the measured linear polarization vectors. Furthermore, we detected a circular 
polarization of 0.6\% toward the brightest \meth ~maser feature (G213.15), which implies a Zeeman splitting of 
$\Delta V_{\rm{Z}}=-6.6\pm1.0$~\ms.
\section{Discussion}
\label{discussion}
%
%\subsection{Relation between $P_{\rm{l}}$ and \tbo}
%
%\begin{figure}[th!]
%\centering
%\includegraphics[width = 9 cm]{TbOm_Pl.eps}
%\caption{Modified version of Fig.~6 of Paper~II. Here, the emerging brightness temperatures (\tbo) is plotted as 
%function of the linear polarization fraction ($P_{\rm{l}}$). The blue and red circles indicate the unsaturated and saturated masers, 
%respectively, detected in several star-forming regions by Surcis et al. (\cite{sur11a,sur12,sur13}) and by us. 
%The red arrows indicate that the \tbo ~values obtained from the adapted FRTM code are lower limits. The red full 
%line is the limit of \tbo ~above which the \meth ~masers are considered saturated ($T_{\rm{b}}\Delta\Omega>2.6\times10^9$~K~sr; Surcis 
%et al. \cite{sur11a}), and the dotted line gives the lower limit to the linear polarization fraction for saturated masers 
%($P_{\rm{l}}\approx4.5~\%$, Paper~I).}
%\label{pltb}
%\end{figure}
%
%%
\begin{table*}[th!]
\caption []{Comparison between the position angle of the magnetic field, \meth ~maser distribution, outflows, and linear polarization angles.} 
\begin{center}
\scriptsize
\begin{tabular}{ l c c c c c c c c c c}
\hline
\hline
\,\,\,\,\,(1) &(2)           & (3)                  & (4)                           & (5)                          & (6)                       & (7)        & (8)                                                   & (9)                                           &(10)                                               &(11)\\
Source & $\Phi_{\rm{f}}$\tablefootmark{a}& $\langle\chi\rangle$\tablefootmark{b} & $\langle\Phi_{\rm{B}}\rangle$\tablefootmark{b} & $\rm{PA}_{\rm{outflow}}$     & $\rm{PA}_{\rm{CH_{3}OH}}$ & $\rho$\tablefootmark{c}& $|\rm{PA}_{\rm{outflow}}-\langle\Phi_{\rm{B}}\rangle|$& $|\rm{PA}_{\rm{CH_{3}OH}}-\langle\chi\rangle|$& $|\rm{PA}_{\rm{CH_{3}OH}}-\rm{PA}_{\rm{outflow}}|$&ref.\tablefootmark{d} \\ 
       & (\d)                &  (\d)                & (\d)                          & (\d)                         & (\d)                      &            & (\d)                                                  &(\d)                                           & (\d)                                              & \\ 
\hline
IRAS\,20126+4104 & $4$       & $-70\pm16$           & $+20\pm16$                    & $-65\pm5$\tablefootmark{e}   & $+87\pm4$                 &  $+0.12$   & $85\pm17$                                             & $23\pm17$\tablefootmark{f}                    & $28\pm6$\tablefootmark{f}                         & (1), (2)\\
G24.78+0.08-A2 &   $17$      & $-53\pm2$            & $+37\pm2$\tablefootmark{g}    & $-40\pm15$\tablefootmark{h}  & $-26\pm19$                &  $-0.77$   & $77\pm15$                                             & $79\pm19$                                     & $66\pm24$                                         & (3), (4)\\
G25.65+1.05    &   $7$       & $-80\pm8$            & $-23\pm51$\tablefootmark{g}   & $-15\pm15$\tablefootmark{h}  & $-49\pm7$\tablefootmark{i}&$-0.87$     & $8\pm53$                                              & $31\pm11$                                     & $64\pm17$                                         & (3), (5)\\
G29.86-0.04    &   $17$      & $+46\pm41$           & $+82\pm56$\tablefootmark{g}   & $+6\pm15$\tablefootmark{h,j} & $+8\pm7$\tablefootmark{i} &$+0.73$     & $76\pm58$                                             & $38\pm42$                                     & $14\pm17$                                         & (3), (6)\\
G35.03+0.35    &   $8$       & $-64\pm5$            & $+26\pm5$\tablefootmark{g}    & $+27\pm15$\tablefootmark{h,k}&$-26\pm19$                 &$-0.77$     & $1\pm16$                                              & $38\pm20$                                     & $53\pm24$                                         & (3), (7), (8)\\
G37.43+1.51    &   $4$       & $+90\pm3$            & $+90\pm3$\tablefootmark{g}    & $-4\pm15$\tablefootmark{h}   &$-64\pm5$\tablefootmark{i} &$-0.87$     & $86\pm15$\tablefootmark{f}                            & $26\pm6$\tablefootmark{f}                     & $60\pm16$                                         & (3), (9)\\
G174.20-0.08   &   $4$       & $-$                  & $-$                           & $-40\pm15$\tablefootmark{h}  &$-63\pm16$                 &$-0.45$     & $-$                                                   & $-$                                           & $23\pm22$                                         & (3), (10)\\
G213.70-12.6-IRS3&$2$        & $+20\pm5$            & $-70\pm5$\tablefootmark{g}    & $+53\pm15$\tablefootmark{h}  &$+63\pm2$                  &$+0.95$     & $57\pm16$\tablefootmark{f}                            & $43\pm5$                                      & $10\pm15$                                         & (3), (11)\\
\hline
\multicolumn{10}{c}{From Paper~II\tablefootmark{l}.}\\
\hline
Cepheus~A      &   $2$       & $-57\pm28$           & $+30\pm19$                    & $+40\pm4$                    & $-79\pm9$                 &  $-0.34$   & $10\pm19$                                             & $22\pm29$                                     & $61\pm10$                                         & (12)\\
W75N-group~A & $3$           & $-13\pm9$            & $+77\pm9$                     & $+66\pm15$                   & $+43\pm10$                &  $+0.96$   & $11\pm18$                                             & $56\pm14$                                     & $23\pm18$                                         & (12)\\
NGC7538-IRS1 &  $6$          &$-30\pm69$            & $+67\pm70$                    & $-40\pm10$                   & $+84\pm7$                 &  $+0.15$   & $73\pm71$                                                 & $66\pm69$                                     & $56\pm12$                                         & (12)\\
W3(OH)-group II &  $4$       &$+21\pm45$            & $-47\pm44$                    & $-$                          & $-59\pm6$                 &  $-0.84$   & $-$                                                   & $80\pm45$                                     & $-$                                               & (12)\\
W51-e2    &  $12$            &$+33\pm16$            & $-60\pm21$                    & $-50\pm20$                   & $+57\pm8$                 &  $+0.70$   & $10\pm29$                                             & $24\pm18$                                     & $73\pm22$                                         & (12)\\
IRAS18556+0138 &  $5$        &$-2\pm11$             & $+88\pm11$                    & $+58\pm23$                   & $-40\pm2$                 &  $-0.99$   & $30\pm26$                                             & $42\pm11$                                     & $82\pm23$                                         & (12)\\
W48       & $7$              & $+23\pm7$            & $-67\pm7$                     & $-$                          & $+55\pm10$                &  $+0.70$   & $-$                                                   & $78\pm12$                                     & $-$                                               & (12) \\
IRAS06058+2138-NIRS1 &  $4$  &$+49\pm47$            & $-49\pm52$                    & $-50\pm15$                   & $+78\pm7$                 &  $+0.64$   & $1\pm54$                                              & $29\pm48$                                     & $52\pm17$                                         & (12)\\
IRAS22272+6358A &  $2$       &$-80\pm15$            & $+9\pm15$                     & $-40\pm15$                   & $-35\pm11$                &  $-0.87$   & $49\pm21$                                             & $45\pm19$                                     & $5\pm19$                                          & (12) \\
S255-IR   &  $4$             &$+36\pm12$            & $-54\pm12$                    & $+75\pm15$                   & $-63\pm49$                &  $-0.11$   &$51\pm19$                                              & $81\pm51$                                     & $42\pm51$                                         & (12) \\
S231      &  $4$             &$+28\pm49$            & $-62\pm49$                    & $-47\pm5$                    & $+28\pm8$                 &  $+0.97$   & $15\pm49$                                             & $0\pm50$                                      & $75\pm9$                                          & (12)\\
G291.27-0.70 &  $7$          &$-32\pm5$             & $+52\pm5$                     & $-$                          & $-77\pm14$                &  $-$       & $-$                                                   & $45\pm15$                                     & $-$                                               & (12)\\
G305.21+0.21 &  $9$          &$-51\pm14$            & $28\pm14$                     & $-$                          & $+48\pm23$                &  $-$       &$-$                                                    & $81\pm27$                                     & $-$                                               & (12)\\
G309.92+0.47 &  $12$         &$+2\pm56$             & $-75\pm56$                    & $-$                          & $+35\pm5$                 &  $-$       &$-$                                                    & $33\pm56$                                     & $-$                                               & (12)\\
G316.64-0.08 &  $3$          &$-67\pm36$            & $+21\pm36$                    & $-$                          & $+34\pm29$                &  $-$       & $-$                                                   & $79\pm46$                                     & $-$                                               & (12)\\
G335.79+0.17 &  $8$          &$+44\pm28$            & $-41\pm28$                    & $-$                          & $-69\pm25$                &  $-$       & $-$                                                   & $67\pm38$                                     & $-$                                               & (12) \\
G339.88-1.26 &  $7$          &$+77\pm24$            & $-12\pm24$                    & $-$                          & $-60\pm17$                &  $-$       & $-$                                                   & $43\pm29$                                     & $-$                                               & (12) \\
G345.01+1.79 &  $5$          &$+5\pm39$             & $-86\pm39$                    & $-$                          & $+74\pm4$                 &  $-$       &$-$                                                    & $69\pm39$                                     & $-$                                               & (12) \\
NGC6334F (central) &  $5$    &$+77\pm20$            & $-13\pm20$                    & $+30\pm15$\tablefootmark{h}  & $-41\pm16$                &  $-$       &$43\pm25$                                              & $62\pm26$                                          & $71\pm41$                                         & (12); (13)\\
NGC6334F (NW)&  $5$          &$-71\pm20$            & $+19\pm20$                    & $+30\pm15$\tablefootmark{h}  & $-80\pm38$                &  $-$       &$11\pm25$                                              & $9\pm43$                                      & $70\pm41$\tablefootmark{f}                        & (12); (13)\\
\hline
\hline
\end{tabular}
\end{center}
\tablefoot{
\tablefoottext{a}{Foreground Faraday rotation estimated by using Eq.~3 of Paper~I.
\tablefoottext{b}{Because of the high uncertainties of the estimated $\Phi_{\rm{f}}$, the angles are not corrected for $\Phi_{\rm{f}}$.}
\tablefoottext{c}{Pearson product-moment correlation coefficient $-1\leq\rho\leq+1$; $\rho=+1$ ( $\rho=-1$) is total positive (negative) correlation, $\rho=0$ is no correlation.}}
\tablefoottext{d}{(1) Surcis et al. (\cite{sur14}); (2) Moscadelli et al. (\cite{mos11}); (3) this work; (4) Beltr\'{a}n et al. (\cite{bel11}); 
(5) S\'{a}nchez-Monge et al. (\cite{san13}); (6) de Villiers et al. (\cite{dev14}); (7) Cyganowski et al. (\cite{cyg09});
(8) Paron et al. (\cite{par12}); (9) L\'{o}pez-Sepulcre et al. (\cite{sep10}); (10) Goddi et al. (\cite{god11}); (11) Dierickx et al. (\cite{die15}); 
(12) Paper~II} and references therein; (13) Zhang et al. (\cite{zha14}).
\tablefoottext{e}{We overestimate the errors by considering half of the opening angle of the outflow.}
\tablefoottext{f}{The differences between the angles are evaluated taking into account that $\rm{PA}\equiv\rm{PA}\pm180$\d,  
$\langle\chi\rangle\equiv\langle\chi\rangle\pm180$\d, and $\langle\Phi_{\rm{B}}\rangle\equiv\langle\Phi_{\rm{B}}\rangle\pm180$\d.}
\tablefoottext{g}{Before averaging, we use the criterion described in Sect.~\ref{obssect} to estimate the orientation of the magnetic field w.r.t the 
linear polarization vectors.}
\tablefoottext{h}{We consider an arbitrary conservative error of 15\d.}
\tablefoottext{i}{We consider only group~A.}
\tablefoottext{j}{We consider the PA of the redshifted lobe of the CO outflow; see Sect.~\ref{Borient}.}
\tablefoottext{k}{We assumed $\rm{PA_{outflow}}=\rm{PA^{4.5\,\mu m}}$; see Sect.~\ref{G35_sec}.}
\tablefoottext{l}{Here we omit all the notes that are already included in Table~2 of Paper~II.}
}
\label{Comp_ang}
\end{table*}
\subsection{Magnetic field orientations}
\label{Borient}
Linear polarization vectors may undergo a rotation when the radiation crosses a medium that is immersed in a magnetic field. This phenomenon is 
known as Faraday rotation. Because the polarized maser emission may be affected by two of these Faraday rotations,
the internal ($\Phi_{\rm{i}}$) and the foreground Faraday rotation ($\Phi_{\rm{f}}$), we  briefly determined whether their effects are
negligible or not. The former, that is, $\Phi_{\rm{i}}$, can be considered negligible as explained in Papers~I and~II, while
$\Phi_{\rm{f}}$ needs to be estimated numerically by using Eq.~3 of Paper~I. We find that $\Phi_{\rm{f}}$ ranges between about 
2\d ~and 17\d, for four sources it is within the errors of the measured linear polarization angles (see 
Tables~\ref{G24_tab}--\ref{G213_tab}), and for three sources it is larger. However, $\Phi_{\rm{f}}$ is very uncertain because the 
errors of some parameters used to calculate it cannot be estimated. Therefore we did not correct either the $\langle\chi\rangle$ angles or the $\langle\Phi_{\rm{B}}\rangle$ angles, but we list $\Phi_{\rm{f}}$ in Col.2 of Table~\ref{Comp_ang} for reader judgment.\\
\indent We now discuss separately the orientation of the magnetic field in the massive SFRs toward which we detected
linearly polarized \meth ~maser emission.\\

\noindent \textit{\textbf{G24.78+0.08.}} Taking into account that for G24.23 and G24.52 the magnetic field is derived to be 
parallel to the linear polarization vector, the error-weighted orientation of the magnetic field around A1 and A2 is 
$\langle\Phi_{\rm{B}}^{\rm{A1}}\rangle=+39^{\circ}\pm42$\d ~and $\langle\Phi_{\rm{B}}^{\rm{A2}}\rangle=+37^{\circ}\pm2$\d, respectively.
Although for G24.78+0.08 $\Phi_{\rm{f}}=17$\d, the magnetic fields in both cores are oriented preferentially along the velocity gradient
of the toroidal structures (PA$_{\rm{A1}}=+50$\d ~and 
PA$_{\rm{A2}}=+40$\d; Beltr\'{a}n et al. \cite{bel11}), and not along the CO-outflow ($\rm{PA_{outflow}^{\rm{^{12}CO}}}=-40$\d; 
Beltr\'{a}n et al. 
\cite{bel11}), indicating that the magnetic field is possibly
located on their surfaces (see Fig.~\ref{G24_cp}). Furthermore, in A1 the 
Zeeman-splitting measurements are spatially distributed with the negative measurement in the northern maser group and the positive measurement 
in the southern maser group (see left panel of Fig.~\ref{G24_cp}).
We recall that
if $\Delta V_{\rm{Z}}>0,$ the magnetic field points away from the observer, and if $\Delta V_{\rm{Z}}<0,$ toward the observer,
the magnetic field around A1 shows a counterclockwise direction that is opposite to the rotation of the toroidal structure. This is
similar to what Surcis et al. (\cite{sur11a}) measured in NGC7538. We also note that from our measurements the magnetic field 
seems to wrap the gas along the preferential southeast-northwest 
direction of star formation (Beltr\'{a}n et al. \cite{bel11}). We measured $\Delta V_{\rm{Z}}<0$ toward A2, but in this 
case, because we have only one measurement, we cannot determine if the magnetic field behaves similarly to the field associated with A1.
 Unfortunately, we cannot discern if the magnetic field is associated 
directly with the two toroidal structures or with the gas that surrounds all the cores. \\

\noindent \textit{\textbf{G25.65+1.05.}} We measured an error-weighted orientation of the magnetic field of 
$\langle\Phi_{\rm{B}}\rangle=-23^{\circ}\pm51$\d ~by taking into account that for G25.02 and G25.06 
$|\theta^{\rm{+}}-55$\d$|<|\theta^{\rm{-}}-55$\d$|$ (see Sect.~\ref{G25_sec}). Therefore, the magnetic field is oriented along 
the SiO outflow ($\rm{PA_{outflow}^{\rm{SiO}}}=-15$\d; S\'{a}nchez-Monge et al. \cite{san13}). For G25.02 
$|\theta^{\rm{+}}-55$\d$|$ is smaller than $|\theta^{\rm{-}}-55$\d$|$ by only 1\d. This indicates that the probability that the 
magnetic field is parallel to the linear polarization vector is not as high as to completely exclude the opposite. However, 
even if we consider that the magnetic field is perpendicular to the linear polarization vector of G25.02, we still have that the 
magnetic field ($\langle\Phi_{\rm{B}}'\rangle=+1^{\circ}\pm37$\d) is preferentially oriented along the outflow. If we now compare 
our measurements, both 
$\langle\Phi_{\rm{B}}\rangle$ and $\langle\Phi_{\rm{B}}'\rangle$, with the measurement of the magnetic field at arcsecond scale 
($\rm{\Phi_{B}^{760\mu m}=+8^{\circ}\pm16}$\d; Vall\'{e}e \& Bastien \cite{val00}), we find a good agreement within the errors.\\
\begin{figure*}[th!]
\centering
\includegraphics[width = 8 cm]{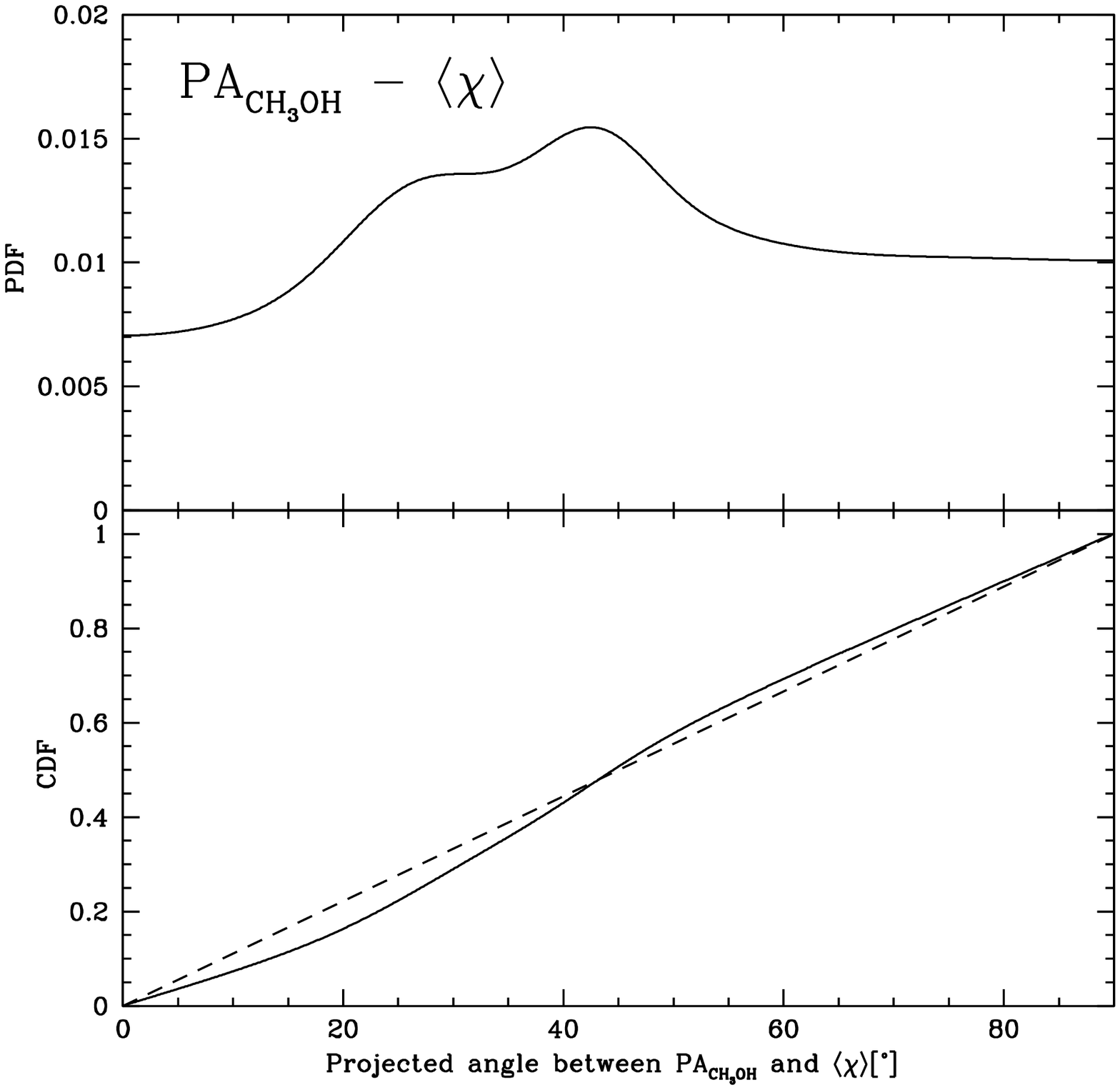}
\includegraphics[width = 8 cm]{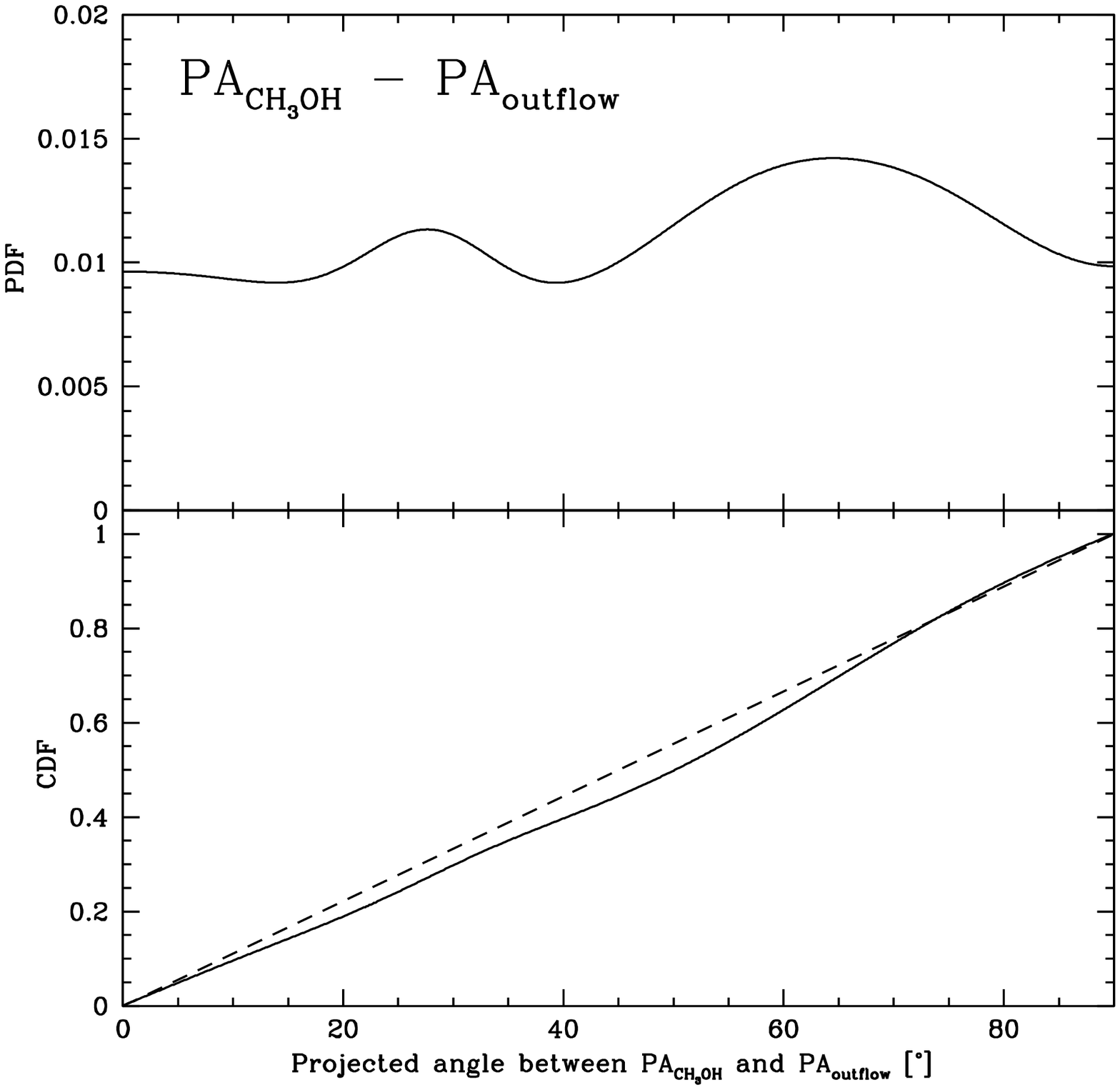}
\caption{Left: The probability distribution function (PDF, top panel) and the cumulative distribution function (CDF, bottom panel) of the
projected angle between the PA of the \meth ~maser distribution and the linear polarization angles 
($|\rm{PA}_{\rm{CH_{3}OH}}-\langle\chi\rangle|$). Right: The PDF and the CDF of the projected angle between the PA of the \meth ~maser
 distribution and the outflow axes ($|\rm{PA}_{\rm{CH_{3}OH}}-\rm{PA_{\rm{outflow}}}|$). In both panels the dashed line is the 
CDF for random orientation of outflows and magnetic fields, i.e., all angular differences are equally likely. The results of the 
Kolmogorov-Smirnov test are listed in Table~\ref{KS}.}
\label{cdf}
\end{figure*}

\noindent \textit{\textbf{G29.86-0.04.}} Although $\Phi_{\rm{f}}=17$\d ~is estimated to be large, the magnetic field is oriented 
almost east-west on the plane of the sky ($\langle\Phi_{\rm{B}}\rangle=+82^{\circ}\pm56$\d), which is consistent with the orientation of 
the bent blueshifted lobe of the CO-outflow. However, because the magnetic field orientation is estimated from the polarized emission 
of masers that are spatially associated with the redshifted lobe of the CO-outflow, we must only compare it with the orientation 
of the redshifted lobe.
Therefore, the magnetic field is almost perpendicular to it, suggesting that perhaps the \meth ~masers probe a magnetic field 
that might be twisted around the axis of the redshifted lobe of the CO-outflow. However, because we have only 
one Zeeman-splitting measurement, which indicates that the magnetic field is pointing toward the observer, our interpretation is
merely speculative.\\

\noindent \textit{\textbf{G35.03+0.35.}} We were able to determine the magnetic field orientation on the plane of the sky 
from one linear polarization measurement ($\Phi_{\rm{B}}=+26^{\circ}\pm5$\d). 
The magnetic field is oriented along the 4.5~$\mu$m emission and the projection on the plane of the sky of the CO-outflow.\\

\noindent \textit{\textbf{G37.43+1.51.}} The magnetic field is assumed to be parallel to all the linear polarization 
vectors of the 6.7 GHz \meth ~masers measured toward group~A of this massive SFR, that is, $\langle\Phi_{\rm{B}}\rangle=+90^{\circ}\pm3$\d. 
The magnetic field is thus perpendicular to the orientation on the plane of the sky of the C$^{18}$O-outflow 
($\rm{PA_{outflow}^{\rm{C^{18}O}}}=-4$\d; L\'{o}pez-Sepulcre et al. \cite{sep10}). \\

\noindent \textit{\textbf{G213.70-12.6.}} The magnetic field at mas resolution shows an orientation on the plane of the sky of 
$\langle\Phi_{\rm{B}}\rangle=-70^{\circ}\pm5$\d. The magnetic field thus appears to be aligned with the linear 
polarization vector of Star~A ($\rm{PA}^{\rm{Star~A}}_{P_{\rm{l}}}=-68^{\circ}\pm2$\d; Simpson et al. \cite{sim13}), which the \meth 
~masers are associated with, but it is rotated by about 90\d ~with respect to the magnetic field inferred\footnote{If the 
magnetic field 
is considered to be perpendicular to the linear polarization vectors of the infrared emissions.} from the polarimetric measurements at 
scales larger than $0''\!.2$ (Yao et al. \cite{yao97}; Simpson et al. \cite{sim13}). The magnetic field probed by the 
masers is almost aligned with the large-scale CO-outflow detected toward IRS\,6 ($\rm{PA}_{\rm{outflow}}^{\rm{CO}}\approx-45$\d; Xu et al. 
\cite{xu06}), but it is almost perpendicular to the small-scale $^{13}$CO(2-1) outflow associated with IRS\,3 
($\rm{PA}_{\rm{outflow}}^{\rm{CO}}\approx+53$\d; Dierickx et al. \cite{die15}). Similarly to G29.86-0.04, we here speculate that the 
magnetic field in G213.70-12.6 might be twisted along the outflow axis. \\
\indent The Zeeman splitting is measured from the circularly polarized spectra of the brightest maser G213.15 (Fig.~\ref{Vfit}).
Because G213.15 is assumed to be partially saturated (see Sect.~\ref{G213_sec}), the circular polarization might be influenced 
by a non-Zeeman effect due to the saturation state of the maser, that is, the rotation of the axis of symmetry for the molecular quantum 
states (e.g., Vlemmings \cite{vle08}). Although this effect is difficult to quantify,
we are quite confident that the contribution to $P_{\rm{V}}$ of this non-Zeeman effect is not high enough to invert the $S$-shape of 
the $V$ spectra. Otherwise, we would have measured a much higher value of \tbo \ for G213.15, which is 
slightly above the saturation threshold of $\rm{log}(T_{\rm{b}}\Delta\Omega)=9.4$~log(K sr). Consequently, from the sign 
of the Zeeman splitting, we can conclude that the magnetic field is pointing toward the observer.
\subsection{Updated statistical results}
\begin{figure}[t!]
\centering
\includegraphics[width = 8 cm]{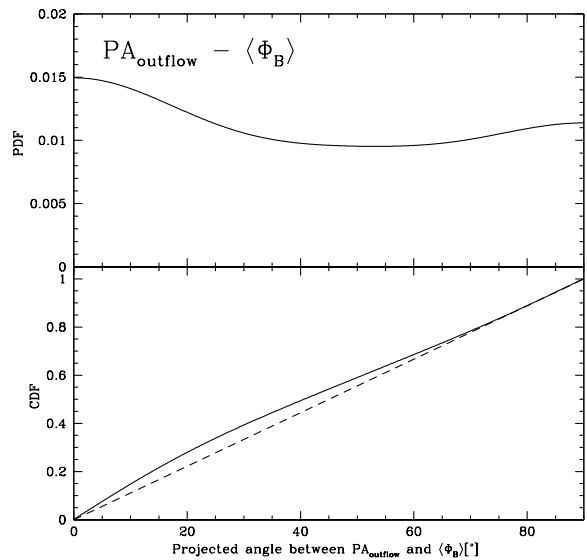}
\caption{Probability distribution function (PDF, top panel) and the cumulative distribution function (CDF, bottom panel) of the
projected angle between the magnetic field and the outflow axes ($|\rm{PA}_{\rm{outflow}}-\langle\Phi_{\rm{B}}\rangle|$). The dashed line 
is the CDF for random orientation of outflows and magnetic fields, i.e., all angular differences are equally likely. The results of the 
Kolmogorov-Smirnov test are listed in Table~\ref{KS}}
\label{cdf2}
\end{figure}

At the midpoint of our project to determine if there exists any relation between the morphology of the magnetic field on a scale 
of tens of astronomical unit and the ejecting direction of molecular outflow from massive YSOs, we must update our first 
statistical results reported in Paper~II by adding the
new magnetic field measurements made around the sources discussed in Sect.~\ref{Borient} and around IRAS\,20126+4104 (Surcis et al. 
\cite{sur14}). Moreover, we also added two of the southern sources observed by Dodson \& Moriarty (\cite{dod12})
to our analysis: NGC6334(central) and NGC6334(NW), which were recently associated with the blueshifted lobe of a CO-outflow (Zhang et al. 
\cite{zha14}). Therefore we analyzed the probability distribution function (PDF) and the cumulative 
distribution function (CDF) of the projected angles $|\rm{PA}_{\rm{outflow}}-\langle\Phi_{\rm{B}}\rangle|$, 
$|\rm{PA}_{\rm{CH_{3}OH}}-\langle\chi\rangle|$, and $|\rm{PA}_{\rm{CH_{3}OH}}-\rm{PA}_{\rm{outflow}}|$; where $\rm{PA}_{\rm{outflow}}$
is the orientation of the large-scale molecular outflow on the plane of the sky, $\langle\Phi_{\rm{B}}\rangle$ is the error -weighted orientation of the magnetic field on the plane of the sky, $\rm{PA}_{\rm{CH_{3}OH}}$ is the orientation of the \meth ~maser 
distribution, and $\langle\chi\rangle$ is the error-weighted value of the linear polarization angles.
Note that although Surcis et al. (\cite{sur14}) determined the morphology of the magnetic field around IRAS\,20126+4104 by 
observing the polarized emission of both 6.7 GHz \meth ~and 22 GHz \water ~masers, we consider here only the orientation of the magnetic 
field estimated from the \meth ~masers.\\
\indent We list all the sources 
of the updated magnetic field total sample in Table~\ref{Comp_ang}; note that all the angles are the projection on the plane of the 
sky. For the statistical analysis we require the uncertainties of all the angles. While the errors of $\rm{PA}_{\rm{CH_{3}OH}}$ 
and $\langle\chi\rangle$ are 
easily determined, the uncertainties of $\rm{PA}_{\rm{outflow}}$ are unknown for all the new sources but IRAS\,20126+4104. Therefore, as 
already done in Paper~II, we considered a conservative uncertainty of $\pm15$\d. The uncertainties in Cols. 8 to 10 of 
Table~\ref{Comp_ang} are 
equal to $\sigma_{\rm{x-y}}=\sqrt{\sigma^{2}_{\rm{x}}+\sigma^{2}_{\rm{y}}}$, where x and y are the two angles taken in consideration 
in each column.\\
\indent In Figs.~\ref{cdf} and~\ref{cdf2} we show the PDF and the CDF of $|\rm{PA}_{\rm{CH_{3}OH}}-\langle\chi\rangle|$, 
$|\rm{PA}_{\rm{CH_{3}OH}}-\rm{PA}_{\rm{outflow}}|$, and
$|\rm{PA}_{\rm{outflow}}-\langle\Phi_{\rm{B}}\rangle|$. The results of the Kolmogorov-Smirnov (K-S) test are reported in 
Table~\ref{KS}. We note that the probability that the angles $|\rm{PA}_{\rm{CH_{3}OH}}-\langle\chi\rangle|$ are drawn from a random 
distribution is now $\sim$80\%, which is 20\% higher than what we computed in Paper~II. On the other hand, the probability for the 
angles $|\rm{PA}_{\rm{CH_{3}OH}}-\rm{PA}_{\rm{outflow}}|$ decreases to 34\%, which was 60\% in Paper~II.
Note that if more than one maser group is detected toward an SFR region, we consider in our analysis the maser group 
that shows the longest linear distribution and that is clearly associated with the outflow. Even for scattered maser distribution 
(e.g., G174.20-0.08 MM1) we perform a linear fit.\\
\indent Although the number of sources for which molecular outflows have been detected and for which the orientation of the magnetic 
field has been determined is now twice that of Paper~II (18 vs. 9), the probability that the distribution
of $|\rm{PA}_{\rm{outflow}}-\langle\Phi_{\rm{B}}\rangle|$ values are drawn from a random distribution is still 10\%. This probability 
confirms our previous conclusion: the magnetic field close to the central YSO (10-100~au) is preferentially oriented 
along the outflow axis.\\
\indent A more accurate statistical analysis will be presented in the last paper of the series when all the sources are
observed and analyzed.
\section{Summary}
\begin {table}[t!]
\caption []{Results of the Kolmogorov-Smirnov test.} 
\begin{center}
\scriptsize
\begin{tabular}{ l c c c c }
\hline
\hline
\,\,\,\,\,\,\,\,\,\,\,\,\,\,\,(1)                      &(2)   & (3)  & (4)       & (5)          \\ 
\,\,\,\,\,\,\,\,\,\,Angle                              & $N$\tablefootmark{a}  & $D$\tablefootmark{b}  & $\lambda$\tablefootmark{c} & $Q_{\rm{K-S}}(\lambda)$\tablefootmark{d}\\
\hline
\\
$|\rm{PA}_{\rm{CH_{3}OH}}-\langle\chi\rangle|$         & 27   & 0.12 & 0.65      & 0.79 \\
$|\rm{PA}_{\rm{CH_{3}OH}}-\rm{PA_{\rm{outflow}}}|$     & 19   & 0.21 & 0.94      & 0.34 \\
$|\rm{PA}_{\rm{outflow}}-\langle\Phi_{\rm{B}}\rangle|$ & 18   & 0.28 & 1.22      & 0.10 \\
\\
\hline
\hline
\end{tabular}
\end{center}
\tablefoot{
\tablefoottext{a}{$N$ is the number of elements considered in the K.-S. test. }
\tablefoottext{b}{$D$ is the highest value of the absolute difference between the data set, composed of $N$ elements, and the random
distribution.}
\tablefoottext{c}{$\lambda$ is a parameter given by $\lambda=(\sqrt{N}+0.12+0.11/\sqrt{N})\times D$.}
\tablefoottext{d}{$Q_{\rm{K-S}}(\lambda)=2\sum_{j=1}^{N} (-1)^{j-1}~ e^{-2j^2\lambda^2}$ is the significance level of the K-S test.}
}
\label{KS}
\end{table}
We observed seven massive star-forming regions at 6.7 GHz in full polarization spectral mode with the EVN to detect the linearly and
circularly polarized emission of \meth ~masers. We detected linearly polarized emission toward all the sources but 
G174.20-0.08 (AFGL\,5142) and circularly polarized emission toward three sources,  G24.78+0.08, G29.86-0.04, and 
G213.70-12.6.
By analyzing the polarized emission of the masers, we were able to estimate the orientation of the magnetic field around seven massive
YSOs, considering that G24.78+0.08 hosts two centers of \meth ~masers around each of the YSOs A1 and A2. The magnetic field is 
oriented along the outflows in two YSOs, it is almost perpendicular to the outflows in four YSOs, and in one
YSOs (G24.78+0.08 A1) a 
comparison is not possible. Moreover, in G24.78+0.08 A1 and A2 the magnetic field is oriented along the toroidal structures. From the 
circularly polarized emission of the \meth ~masers we measured Zeeman splitting toward G24.78+0.08 (both A1 and A2), G29.86-0.04, 
and G213.70-12.6.\\
\indent We added all the magnetic field measurements made toward the YSOs presented in this work to the magnetic field total 
sample, which contains all the massive YSOs observed so far in full polarization mode at 6.7 GHz anywhere on the sky. Similarly to Paper~II, we compared the projected angles between magnetic fields and outflows. We still find evidence that 
the magnetic field around massive YSOs are preferentially oriented along the molecular outflows. Indeed,
the Kolmogorov-Smirnov test still shows a probability of 10\% that our distribution of angles is drawn from a random 
distribution.\\

\noindent \small{\textit{Acknowledgements.} We wish to thank the anonymous referee for useful suggestions that have improved 
the paper. W.H.T.V. acknowledges support from the European Research Council through consolidator grant 614264. 
A.B. acknowledges support from the National Science Centre Poland through grant 2011/03/B/ST9/00627. 
M.G.B. acknowledges the JIVE Summer Student Programme 2013.}

\bibliographystyle{aa}

\begin{thebibliography}{}
\small
\bibitem[2000]{arg00}
Argon, A.L., Reid, M.J., \& Menten, K.M. 2000, ApJS, 129, 159
\bibitem[2004]{bel04}
Beltr\'{a}n, M.T., Cesaroni, R., Neri, R. et al., 2004, ApJ, 601, L187
\bibitem[2005]{bel05}
Beltr\'{a}n, M.T., Cesaroni, R., Neri, R. et al., 2005, A\&A, 435, 901
\bibitem[2006]{bel06}
Beltr\'{a}n, M.T., Cesaroni, R., Codella, C. et al., 2006, Nature, 443, 427
\bibitem[2011]{bel11}
Beltr\'{a}n, M.T., Cesaroni, R., Zhang, Q. et al. 2011, A\&A, 532, A91
\bibitem[2011]{bre11}
Breen, S.L. \& Ellingsen, S.P. 2011, MNRAS, 416, 178
\bibitem[1997]{car97}
Carpenter, J.M., Meyer, M.R., Dougados, M.R. et al. 1997, AJ, 114, 198
\bibitem[1993]{cas93}
Caswell, J.L., Gardner, F.F., Norris, R.P. et al. 1993, MNRAS, 260, 425
\bibitem[1995]{cas95}
Caswell, J.L., Vaile, R.A., Ellingsen, S.P. et al. 1995, MNRAS, 272, 96
\bibitem[2013]{cas13}
Caswell, J.L., Green, J.A., \& Phillips, C.J. 2013, MNRAS, 431, 1180
\bibitem[2003]{ces03}
Cesaroni, R., Codella, C., Furuya, R.S. et al. 2003, A\&A, 401, 227
\bibitem[2013]{cha13}
Chapman, N.L., Davidson, J.A., Goldsmith, P.F. et al. 2013, ApJ, 770, 151
\bibitem[1997]{cod97}
Codella, C., Testi, L. \& Cesaroni, R. 1997, A\&A, 325, 282
\bibitem[2013]{cod13}
Codella, C., Beltr\'{a}n, M.T., Cesaroni, R. et al. 2013, A\&A, 550, A81
\bibitem[2009]{cyg09}
Cyganowski, C.J., Brogan, C.L., Hunter, T.R. et al. 2009, ApJ, 702, 1615 
\bibitem[2011]{cyg11}
Cyganowski, C.J., Brogan, C.L., Hunter, T.R. et al. 2011, ApJ, 743, 56 
\bibitem[2007]{cur07}
Curran, R.L. \& Chrysostomou, A. 2007, MNRAS, 382, 699
\bibitem[2014]{dev14}
de Villiers, H.M., Chrysostomou, A., Thompson, M.A. et al. 2014, MNRAS, 444, 566
\bibitem[2015]{die15}
Dierickx, M., Jim\'{e}nez-Serra, I., Rivilla, V.M. et al. 2015, \textit{arXiv:1503.05230v1}
\bibitem[2012]{dod12}
Dodson, R. \& Moriarty, C.D. 2012, MNRAS, 421, 2395
\bibitem[1999]{for99}
Forster, J.R. \& Caswell, J.L. 1999, A\&AS, 137, 43
\bibitem[2014]{fuj14}
Fujisawa, K., Sugiyama, K., Motogi, K. et al. 2014, PASJ, 66, 31
\bibitem[2002]{fur02}
Furuya, R.S., Cesaroni, R., Codella, C. et al. 2002, A\&A, 390, L1
\bibitem[2008]{gal08}
Galv\'{a}n-Madrid, R., Rodr\'{i}guez, L.F., Ho, P.T.P. et al. 2008, ApJ, 674, L33
\bibitem[2007]{god07}
Goddi, C., Moscadelli, L., Sanna, A. et al. 2007, A\&A, 461, 1027
\bibitem[2011]{god11}
Goddi, C., Moscadelli, L, Sanna, A. 2011, A\&A, 535, L8
\bibitem[1973]{gol73}
Goldreich, P., Keeley, D.A., \& Kwan, J.Y., 1973, ApJ, 179, 111
\bibitem[1992]{hen92}
Henning, Th., Chini, R. \& Pfau, W. 1992, A\&A, 263, 285
\bibitem[1976]{her76}
Herbst, W. \& Racine, R. 1976, AJ, 81, 840
\bibitem[2005]{hil05}
Hill, T., Burton, M.G., Minier, V. et al. 2005, MNRAS, 363, 405
\bibitem[2006]{hil06}
Hill, T., Thompson, M.A., Burton, M.G. et al. 2006, MNRAS, 368, 1223
\bibitem[1994]{how94}
Howard, E.M., Pipher, J.L., \& Forrest, W.J. 1994, ApJ, 425, 707
\bibitem[2013]{hul13}
Hull, C.L.H., Plambeck, R.L., Bolatto, A.D. et al. 2013, ApJ, 768, 159
\bibitem[1995]{hun95}
Hunter, T.R., Testi, L., Taylor, G.B. et al. 1995, A\&A, 302, 249
\bibitem[2000]{hun00}
Hunter, T.R., Churchwell, E., Watson, C. et al. 2000, ApJ, 119, 2711
\bibitem[2015]{kei14}
Keimpema, A., Kettenis, M.M., Pogrebenko, S.V., et al. 2015, Experimental Astronomy, \textit{arXiv:1502.00467}
\bibitem[1994]{kur94}
Kurtz, S., Churchwell, E., \& Wood, D.O.S. 1994, ApJSS, 91, 659
\bibitem[2010]{sep10}
L\'{o}pez-Sepulcre, A., Cesaroni, R. \& Walmsley, C.M. 2010, A\&A, 517, A66
\bibitem[2003]{mck03}
McKee, C.F. \& Tan, J.C. 2003, ApJ, 585, 850
\bibitem[2007]{mck07}
 McKee, C. F. \& Ostriker E.C. 2007, ARA\&A, 45, 565
\bibitem[2000]{min00}
Minier, V., Booth, R.S., \& Conway, J.E. 2000, A\&A, 362, 1093
\bibitem[1996]{mol96}
Molinari, A., Brand, J., Cesaroni, R. et al. 1996, A\&A, 308, 573
\bibitem[2007]{mos07}
Moscadelli, L., Goddi, C., Cesaroni, R. et al. 2007, A\&A, 472, 867
\bibitem[2011]{mos11}
Moscadelli, L., Cesaroni, R., Riojia, M.J., et al. 2011, A\&A, 526, A66
\bibitem[2013]{mye13}
Myers, A.T., McKee, C.F., Cunningham, A.J. et al. 2013, ApJ, 766, 97
\bibitem[2011]{pal11}
Palau, A., Fuente, A., Girart, J.M. et al. 2011, ApJ, 743, L32
\bibitem[2011]{pan11}
Pandian, J.D., Momjian, E., Xu, Y. et al. 2011, ApJ, 730, 55
\bibitem[2012]{par12}
Paron, S., Ortega, M.E., Petriella, A. et al. 2012, MNRAS, 419, 2206
\bibitem[2005]{pes05}
Pestalozzi, M.R., Minier, V., \& Booth, R.S. 2005, A\&A, 432, 737
\bibitem[2011]{pet11}
Peters, T., Banerjee, R., Klessen, R.S. et al. 2011, ApJ, 729, 72
\bibitem[2002]{pre02}
Preibisch, T., Balega, Y.Y., Schertl, D. et al. 2002, A\&A, 392, 945
\bibitem[2013]{san13}
S\'{a}nchez-Monge, A., L\'{o}pez-Sepulcre, A., Cesaroni, R. et al. 2013, A\&A, 557, A94
\bibitem[2012]{sau12}
Sault, R.J. 2012, EVLA Memo 159
\bibitem[1993]{sch93}
Schutte, A.J., van der Walt, D.J., Gaylard, M.J. et al. 1993, MNRAS, 261, 783
\bibitem[2012]{sei12a}
Seifried, D., Pudritz, R.E., Banerjee, R. et al. 2012, MNRAS, 422, 347
\bibitem[1996]{she96}
Shepherd, D.S. \& Churchwell, E. 1996, ApJ, 457, 267
\bibitem[2013]{sim13}
Simpson, J.P., Whitney, B.A., Hines, D.C. et al. 2013, MNRAS, 435, 3419
\bibitem[1988]{sne88}
Snell, R.L., Huang, Y.-L., Dickman, R.L. et al. 1988, ApJ, 325, 853
\bibitem[2009]{sur09}
Surcis, G., Vlemmings, W.H.T., Dodson, R. et al. 2009, A\&A, 506, 757
\bibitem[2011a]{sur11a}
Surcis, G., Vlemmings, W.H.T., Torres, R.M. et al. 2011a, A\&A, 533, A47
\bibitem[2011b]{sur11b}
Surcis, G., Vlemmings, W.H.T., Curiel, S. et al. 2011b, A\&A, 527, A48
\bibitem[2012]{sur12}
Surcis, G., Vlemmings, W.H.T., van Langevelde, H.J. et al. 2012, A\&A, 541, A47, Paper~I
\bibitem[2013]{sur13}
Surcis, G., Vlemmings, W.H.T., van Langevelde, H.J. et al. 2013, A\&A, 556, A73, Paper~II
\bibitem[2014]{sur14}
Surcis, G., Vlemmings, W.H.T., van Langevelde, H.J. et al. 2014, A\&A, 563, A30
\bibitem[2000]{szy00}
Szymczak, M., Hrynek, G., \& Kus, A.J., 2000, A\&AS, 143, 269
\bibitem[1992]{tor92}
Torrelles, J.M., G\'{o}mez, J.F., Anglada, G. et al. 1992, ApJ, 392, 616
\bibitem[2000]{val00}
Vall\'{e}e, J.P. \& Bastien, P. 2000, ApJ, 530, 806
\bibitem[2010]{var10}
Varricatt, W.P., Davis, C.J., Ramsay, S. et al. 2010, MNRAS, 404, 661
\bibitem[2008]{vle08}
Vlemmings, W.H.T. 2008, A\&A, 484, 773
\bibitem[2010]{vle10}
Vlemmings, W.H.T., Surcis, G., Torstensson, K.J.E. et al. 2010, MNRAS, 404, 134 
\bibitem[2011]{vle11}
Vlemmings, W.H.T., Torres, R.M., \& Dodson, R. 2011, A\&A, 529, A95 
\bibitem[1998]{wal98}
Walsh, A.J., Burton, M.G., Hyland, A.R. et al. 1998, MNRAS, 301, 640
\bibitem[2003]{wal03}
Walsh, A.J., MacDonald, G.H., Alvey, N.D.S et al. 2003, A\&A, 410, 597
\bibitem[1974]{war74}
Wardle, J.F.C. \& Kronberg, P.P. 1974, ApJ, 194, 249
\bibitem[2014]{wu14}
Wu, Y.W., Sato, M., Reid, M.J. et al. 2014, A\&A, 566, A17
\bibitem[2006]{xu06}
Xu, Y., Shen, Z.-Q., Yang, J. et al. 2006, ApJ, 132, 20
\bibitem[1997]{yao97}
Yao, Y., Hirata, N., Ischii, M. et al. 1997, ApJ, 490, 281
\bibitem[2002]{zav02}
Zavagno, A., Deharveng, L., Nadeau, D. et al. 2002, A\&A, 394,225
\bibitem[2007]{zha07}
Zhang, Q., Hunter, T.R., Beuther, H. et al. 2007, ApJ, 658, 1152
\bibitem[2014]{zha14}
Zhang, Q., Qiu, K., Girart, J.M. et al. 2014, ApJ, 792, 116

\end{thebibliography}

\Online
\begin{appendix}
\normalsize
\section{Tables}
\label{appA}
In Tables~\ref{G24_tab}--\ref{G213_tab} we list the parameters of all the \meth ~maser features detected toward the seven massive 
star-forming regions observed with the EVN. The tables are organized as follows. The name of the feature is reported in Col.~1 and 
the group to which they belong in Col.~2. The positions, Cols.~3 and 4, refer to the maser feature used for self-calibration. 
The peak flux density, the LSR velocity ($V_{\rm{lsr}}$), 
and the FWHM ($\Delta v\rm{_{L}}$) of the total intensity spectra of the maser features are reported in Cols.~5 to 7. The peak flux density, $V_{\rm{lsr}}$, and $\Delta v\rm{_{L}}$ are obtained using a Gaussian fit. The mean linear polarization fraction 
($P_{\rm{l}}$) and the mean linear polarization angles ($\chi$) that are measured across the spectrum are reported in Cols.~8 
and 9. The best-fitting 
results obtained by using a model based on the radiative transfer theory of methanol masers for $\Gamma+\Gamma_{\nu}=1~\rm{s^{-1}}$ 
(Vlemmings et al. \cite{vle10}, Surcis et al. \cite{sur11b}) are reported in Cols.~10 (the intrinsic thermal linewidth) and 11 
(the emerging brightness temperature). The errors were determined by analyzing the full probability distribution function. The angle 
between the magnetic field and the maser propagation direction ($\theta$, Col.~14) is determined by using the observed 
$P_{\rm{l}}$ and the fit emerging brightness temperature. The
errors for $\theta$ were also determined by analyzing the full 
probability distribution function. The value of $\theta$ in bold indicates that $|\theta^{\rm{+}}-55$\d$|<|\theta^{\rm{-}}-55$\d$|$, 
that is, the magnetic field is assumed to be parallel to the linear polarization vector (see Sect.~\ref{obssect}). The circular 
polarization fraction ($P_{\rm{V}}$) and the Zeeman splitting ($\Delta V_{\rm{Z}}$) are listed in Cols.~12 and 13. The 
Zeeman splitting is determined by fitting the V Stokes spectra by using the best-fitting results ($\Delta V_{\rm{i}}$ and 
$T_{\rm{b}}\Delta\Omega$).

\begin {table*}[t!]
\caption []{Parameters of the 6.7 GHz \meth ~maser features detected in G24.78+0.08.} 
\begin{center}
\scriptsize
\begin{tabular}{ l c c c c c c c c c c c c c }
\hline
\hline
\,\,\,\,\,(1)&(2)   & (3)                & (4)                & (5)       & (6)           & (7)                 & (8)         & (9)       & (10)                & (11)                     & (12)         &(13)                 & (14)         \\
Maser     & Group   & RA\tablefootmark{a}&Dec\tablefootmark{a}& Peak flux & $V_{\rm{lsr}}$& $\Delta v\rm{_{L}}$ &$P_{\rm{l}}$ &  $\chi$   & $\Delta V_{\rm{i}}$ & $T_{\rm{b}}\Delta\Omega$ & $P_{\rm{V}}$ & $\Delta V_{\rm{Z}}$ & $\theta$\\
          &         &offset              &  offset            & density   &               &                     &             &             &                     &                          &              &                     &      \\ 
          &         & (mas)              &  (mas)             & (Jy/beam) &  (km/s)       &      (km/s)         & (\%)        &   (\d)    & (km/s)              & (log K sr)               &   ($\%$)     &  (m/s)              &(\d)       \\ 
\hline
G24.01    &   A2    &   -1542.297        &     746.773        & $0.092\pm0.005$ &  108.57 &      $0.25$         & $-$         & $-$       & $-$                 & $-$                      & $-$         & $-$                 &$-$ \\ 
G24.02    &   A2    &   -1449.196        &     553.302        & $0.222\pm0.004$ &  106.59 &      $0.28$         & $-$         & $-$       & $-$                 & $-$                      & $-$         & $-$                 &$-$ \\ 
G24.03    &   A2    &   -1416.923        &     510.347        & $0.031\pm0.004$ &  105.41 &      $0.20$         & $-$         & $-$       & $-$                 & $-$                      & $-$         & $-$                 &$-$ \\ 
G24.04    &   A2    &   -1410.906        &     1117.249       & $0.115\pm0.014$ &  113.18 &      $0.47$         & $-$         & $-$       & $-$                 & $-$                      & $-$         & $-$                 &$-$ \\
\hline
\end{tabular} \end{center}
\tablefoot{
\tablefoottext{a}{The reference position is $\alpha_{2000}=+18^{\rm{h}}36^{\rm{m}}12^{\rm{s}}\!.563$ and 
$\delta_{2000}=-07^{\circ}12'10''\!\!.787$ (see Sect.~\ref{obssect}).}
}
\label{G24_tab}
\end{table*}

\begin {table*}[h!]
\caption []{Parameters of the 6.7 GHz \meth ~maser features detected in G25.65+1.05.} 
\begin{center}
\scriptsize
\begin{tabular}{ l c c c c c c c c c c c c c}
\hline
\hline
\,\,\,\,\,(1)&(2)   & (3)               & (4)                & (5)             & (6)           & (7)                 & (8)         & (9)        & (10)                & (11)                    & (12)         &(13)                 &(14)      \\
Maser     & Group   &RA\tablefootmark{a}&Dec\tablefootmark{a}& Peak flux       & $V_{\rm{lsr}}$& $\Delta v\rm{_{L}}$ &$P_{\rm{l}}$ & $\chi$     & $\Delta V_{\rm{i}}$ & $T_{\rm{b}}\Delta\Omega$& $P_{\rm{V}}$ & $\Delta V_{\rm{Z}}$ &$\theta$\\
          &         & offset            &  offset            & density         &               &                     &             &              &                     &                         &              &                     &      \\ 
          &         & (mas)             &  (mas)             & (Jy/beam)       &  (km/s)       &      (km/s)         & (\%)        &   (\d)     & (km/s)              & (log K sr)              &   ($\%$)     &  (m/s)              &(\d)       \\ 
\hline
G25.01    &   A     &  -19.690          &   15.911           & $0.338\pm0.014$ &   41.56       &      $0.11$         & $-$         & $-$        & $-$                 & $-$                     & $-$          & $-$                 &  $-$   \\ 
G25.02    &   A     &  0                &   0                & $6.924\pm0.124$ &   41.78       &      $0.31$         & $0.4\pm0.2$ & $-76\pm11$ & $1.6^{+0.1}_{-0.2}$ & $8.2^{+0.7}_{-1.6}$     & $-$          & $-$                 &$\mathbf{72^{+12}_{-45}}$ \\ 
G25.03    &   A     &  6.886            &  -0.382            & $3.860\pm0.126$ &   42.00       &      $0.17$         & $-$         & $-$        & $-$                 & $-$                     & $-$          & $-$                 &  $-$   \\
\hline
\end{tabular} \end{center}
\tablefoot{
\tablefoottext{a}{The reference position is $\alpha_{2000}=+18^{\rm{h}}34^{\rm{m}}20^{\rm{s}}\!.900$ and 
$\delta_{2000}=-05^{\circ}59'42''\!\!.098$ (see Sect.~\ref{obssect}).}
}
\label{G25_tab}
\end{table*}
\begin {table*}[t!]
\caption []{Parameters of the 6.7 GHz \meth ~maser features detected in G29.86-0.04.} 
\begin{center}
\scriptsize
\begin{tabular}{ l c c c c c c c c c c c c c}
\hline
\hline
\,\,\,\,\,(1)&(2)   & (3)                & (4)                 & (5)             & (6)            & (7)                 & (8)           & (9)       & (10)                & (11)                     & (12)         &(13)                 &(14)      \\
Maser     & Group   & RA\tablefootmark{a}& Dec\tablefootmark{a}& Peak flux       & $V_{\rm{lsr}}$ & $\Delta v\rm{_{L}}$ &$P_{\rm{l}}$   &  $\chi$   & $\Delta V_{\rm{i}}$ & $T_{\rm{b}}\Delta\Omega$ & $P_{\rm{V}}$ & $\Delta V_{\rm{Z}}$ &$\theta$\\
          &         & offset             & offset              & density         &                &                     &               &     &                     &                          &              &                     &      \\ 
          &         & (mas)              & (mas)               & (Jy/beam)       &  (km/s)        &      (km/s)         & (\%)          &   (\d)    & (km/s)              & (log K sr)               &   ($\%$)     &  (m/s)              &(\d)       \\ 
\hline
G29.01    &  B      & -188.953           & -153.172            & $0.150\pm0.003$ & 103.68         &      $0.24$         & $-$           & $-$       & $-$                 & $-$                      & $-$         & $-$                 &$-$ \\ 
G29.02    &  B      & -183.009           & -160.760            & $0.971\pm0.015$ & 102.05         &      $0.29$         & $-$           & $-$       & $-$                 & $-$                      & $-$         & $-$                 &$-$ \\ 
G29.03    &  B      & -157.861           & -185.738            & $0.275\pm0.003$ & 103.85         &      $0.32$         & $-$           & $-$       & $-$                 & $-$                      & $-$         & $-$                 &$-$ \\
\hline
\end{tabular} \end{center}
\tablefoot{
\tablefoottext{a}{The reference position is $\alpha_{2000}=+18^{\rm{h}}45^{\rm{m}}59.572^{\rm{s}}$ and 
$\delta_{2000}=-02^{\circ}45'01''\!\!.573$ (see Sect.~\ref{obssect}).}
\tablefoottext{b}{Because of the degree of saturation, $T_{\rm{b}}\Delta\Omega$ is underestimated, $\Delta V_{\rm{i}}$ 
and $\theta$ are overestimated.}
}
\label{G29_tab}
\end{table*}
\begin {table*}[t!]
\caption []{Parameters of the 6.7 GHz \meth ~maser features detected in G35.03+0.35.} 
\begin{center}
\scriptsize
\begin{tabular}{ l c c c c c c c c c c c c c}
\hline
\hline
\,\,\,\,\,(1)&(2)   & (3)                 & (4)                  & (5)             & (6)            & (7)                 & (8)         & (9)       & (10)                 & (11)                     & (12)         &(13)                 & (14)          \\
Maser     & Group   & RA\tablefootmark{a} & Dec\tablefootmark{a} & Peak flux       & $V_{\rm{lsr}}$ & $\Delta v\rm{_{L}}$ &$P_{\rm{l}}$ &  $\chi$   & $\Delta V_{\rm{i}}$  & $T_{\rm{b}}\Delta\Omega$ & $P_{\rm{V}}$ & $\Delta V_{\rm{Z}}$ &$\theta$\\
          &         &  offset             & offset               & density         &                &                     &             &     &                      &                          &              &                     &      \\ 
          &         &  (mas)              & (mas)                & (Jy/beam)       &  (km/s)        &      (km/s)         & (\%)        &   (\d)    & (km/s)               & (log K sr)               &   ($\%$)     &  (m/s)              &(\d)       \\ 
\hline
G35.01    &         &  78.111             &  64.224              & $0.112\pm0.005$ &  44.88         &      $0.24$         & $-$         & $-$       &  $-$                 & $-$                      & $-$          & $-$                &$-$ \\ 
G35.02    &         &  67.948             & -245.222             & $0.061\pm0.006$ &  46.63         &      $0.22$         & $-$         & $-$       &  $-$                 & $-$                      & $-$         & $-$                 &$-$ \\ 
G35.03    &         &  40.077             &  74.663              & $0.026\pm0.005$ &  45.10         &      $0.23$         & $-$         & $-$       &  $-$                 & $-$                      & $-$         & $-$                 &$-$ \\
\hline
\end{tabular} \end{center}
\tablefoot{
\tablefoottext{a}{The reference position is $\alpha_{2000}=+18^{\rm{h}}54^{\rm{m}}00^{\rm{s}}\!.660$ and 
$\delta_{2000}=+02^{\circ}01'18''\!\!.551$ (see Sect.~\ref{obssect}).}
}
\label{G35_tab}
\end{table*}
\begin {table*}[t!]
\caption []{Parameters of the 6.7 GHz \meth ~maser features detected in G37.43+1.51.} 
\begin{center}
\scriptsize
\begin{tabular}{ l c c c c c c c c c c c c c}
\hline
\hline
\,\,\,\,\,(1)&(2)   & (3)               & (4)                & (5)             & (6)           & (7)                 & (8)         & (9)        & (10)                & (11)                     & (12)         &(13)                  &(14)      \\
Maser     & Group   &RA\tablefootmark{a}&Dec\tablefootmark{a}& Peak flux       & $V_{\rm{lsr}}$& $\Delta v\rm{_{L}}$ &$P_{\rm{l}}$ & $\chi$     & $\Delta V_{\rm{i}}$ & $T_{\rm{b}}\Delta\Omega$ & $P_{\rm{V}}$ & $\Delta V_{\rm{Z}}$  &$\theta$\\
          &         & offset            &  offset            & density         &               &                     &             &              &                     &                          &              &                      &      \\ 
          &         & (mas)             &  (mas)             & (Jy/beam)       &  (km/s)       &      (km/s)         & (\%)        &   (\d)     & (km/s)              & (log K sr)               &   ($\%$)     &  (m/s)               &(\d)       \\ 
\hline
G37.01    &   A     &  -68.621          &   18.314           & $0.147\pm0.004$ &   40.65       &      $0.21$         & $-$         & $-$        & $-$                 & $-$                      & $-$          & $-$                  & $-$       \\ 
G37.02    &   A     &  -56.601          &   15.614           & $0.522\pm0.006$ &   40.79       &      $0.20$         & $-$         & $-$        & $-$                 & $-$                      & $-$          & $-$                  & $-$       \\ 
G37.03    &   A     &  -25.050          &   -2.205           & $0.525\pm0.004$ &   42.23       &      $0.21$         & $-$         & $-$        & $-$                 & $-$                      & $-$          & $-$                  & $-$       \\
\hline
\end{tabular} \end{center}
\tablefoot{
\tablefoottext{a}{The reference position is $\alpha_{2000}=+18^{\rm{h}}54^{\rm{m}}14^{\rm{s}}\!.229$ and 
$\delta_{2000}=+04^{\circ}41'41''\!\!.138$ (see Sect.~\ref{obssect}).}
}
\label{G37_tab}
\end{table*}
\begin {table*}[t!]
\caption []{Parameters of the 6.7 GHz \meth ~maser features detected in G174.20-0.08.} 
\begin{center}
\scriptsize
\begin{tabular}{ l c c c c c c c c c c c c c}
\hline
\hline
\,\,\,\,\,(1)&(2)   & (3)                & (4)                & (5)             & (6)            & (7)                 & (8)         & (9)    & (10)                & (11)                     & (12)         &(13)                & (14)           \\
Maser     & Group   & RA\tablefootmark{a}&Dec\tablefootmark{a}& Peak flux       & $V_{\rm{lsr}}$ & $\Delta v\rm{_{L}}$ &$P_{\rm{l}}$ & $\chi$ & $\Delta V_{\rm{i}}$ & $T_{\rm{b}}\Delta\Omega$ & $P_{\rm{V}}$ & $\Delta V_{\rm{Z}}$&$\theta$\\
          &         &  offset            &  offset            & density         &                &                     &             &          &                     &                          &              &                    &      \\ 
          &         &  (mas)             &  (mas)             & (Jy/beam)       &  (km/s)        &      (km/s)         & (\%)        &   (\d) & (km/s)              & (log K sr)               &   ($\%$)     &  (m/s)             &(\d)       \\ 
\hline
G174.01   &         & -30.813            & -17.113            & $0.047\pm0.005$ & 0.43           &      $0.30$         & $-$         & $-$    &  $-$                & $-$                      & $-$            & $-$                &$-$ \\ 
G174.02   &         & -25.677            & -11.940            & $0.047\pm0.007$ & 1.49           &      $0.13$         & $-$         & $-$    &  $-$                & $-$                      & $-$            & $-$                &$-$ \\ 
G174.03   &         & -20.827            & -49.076            & $0.057\pm0.005$ & 1.88           &      $0.31$         & $-$         & $-$    &  $-$                & $-$                      & $-$            & $-$                &$-$ \\ 
G174.04   &         & -16.548            & -9.541             & $0.095\pm0.008$ & 1.66           &      $0.21$         & $-$         & $-$    &  $-$                & $-$                      & $-$            & $-$                &$-$ \\
\hline
\end{tabular} \end{center}
\tablefoot{
\tablefoottext{a}{The reference position is $\alpha_{2000}=+05^{\rm{h}}30^{\rm{m}}48^{\rm{s}}\!.020$ and 
$\delta_{2000}=+33^{\circ}47'54''\!\!.611$ (see Sect.~\ref{obssect}).}
}
\label{G174_tab}
\end{table*}
\begin {table*}[t!]
\caption []{Parameters of the 6.7 GHz \meth ~maser features detected in G213.70-12.6 (Mon~R2--IRS\,3/Star~A).} 
\begin{center}
\scriptsize
\begin{tabular}{ l c c c c c c c c c c c c c}
\hline
\hline
\,\,\,\,\,(1)&(2)    & (3)               & (4)                & (5)             & (6)            & (7)                 & (8)         & (9)       & (10)                & (11)                     & (12)         &(13)                  &(14)           \\
Maser     &Group     &RA\tablefootmark{a}&Dec\tablefootmark{a}& Peak flux       & $V_{\rm{lsr}}$ & $\Delta v\rm{_{L}}$ &$P_{\rm{l}}$ &   $\chi$  & $\Delta V_{\rm{i}}$ & $T_{\rm{b}}\Delta\Omega$ & $P_{\rm{V}}$ & $\Delta V_{\rm{Z}}$  &$\theta$\\
          &          &offset             & offset             & density         &                &                     &             &             &                     &                          &              &                      &      \\ 
          &          &(mas)              & (mas)              & (Jy/beam)       &  (km/s)        &      (km/s)         & (\%)        &   (\d)    & (km/s)              & (log K sr)               &   ($\%$)     &  (m/s)               &(\d)       \\ 
\hline
G213.01   &          &-444.918           &-209.568            & $0.111\pm0.006$ &    10.11       &   0.17              & $-$         & $-$       &        $-$          &          $-$             & $-$            & $-$                  &    $-$   \\ 
G213.02   &          &-436.616           &-206.261            & $0.098\pm0.005$ &    10.07       &   0.33              & $-$         & $-$       &        $-$          &          $-$             & $-$            & $-$                  &    $-$   \\ 
G213.03   &          &-334.883           &-188.515            & $0.105\pm0.005$ &     9.98       &   0.15              & $-$         & $-$       &        $-$          &          $-$             & $-$            & $-$                  &    $-$   \\
\hline
\end{tabular} \end{center}
\tablefoot{
\tablefoottext{a}{The reference position is $\alpha_{2000}=+06^{\rm{h}}07^{\rm{m}}47^{\rm{s}}\!.860$ and 
$\delta_{2000}=-06^{\circ}22'56''\!\!.626$ (see Sect.~\ref{obssect}).}
\tablefoottext{b}{Because of the degree of the saturation $T_{\rm{b}}\Delta\Omega$ is underestimated, $\Delta V_{\rm{i}}$ 
and $\theta$ are overestimated.}
}
\label{G213_tab}
\end{table*}

\end{appendix}

\end{document}